\documentclass[12pt]{article}
\usepackage{a4wide}
\usepackage{amssymb}
\usepackage{bm}
\begin{document}
{\renewcommand{\thefootnote}{\fnsymbol{footnote}}
\begin{center}
{\LARGE Moments and saturation properties of eigenstates}\\
\vspace{1.5em}
Martin Bojowald,
Jonathan Guglielmon and Martijn van Kuppeveld
\\
\vspace{0.5em}
Department of Physics,
The Pennsylvania State
University,\\
104 Davey Lab, University Park, PA 16802, USA\\
\vspace{1.5em}
\end{center}
}

\setcounter{footnote}{0}

\begin{abstract}
  Eigenvalues are defined for any element of an algebra of observables and do
  not require a representation in terms of wave functions or density
  matrices. A systematic algebraic derivation based on moments is presented
  here for the harmonic oscillator, together with a perturbative treatment of
  anharmonic systems. In this process, a collection of inequalities is
  uncovered which amount to uncertainty relations for higher-order moments
  saturated by the harmonic-oscillator excited states. Similar saturation
  properties hold for anharmonic systems order by order in perturbation
  theory. The new method, based on recurrence relations for moments of a state
  combined with positivity conditions, is therefore able to show new physical
  features.
\end{abstract}

\section{Introduction}

The usual derivation of eigenvalues in model systems of quantum mechanics
seems to suggest that spectral properties are a direct consequence of boundary
conditions imposed on wave functions. However, boundary conditions are a
property of representations of an algebra of observables ${\cal A}$ (with a
unit ${\mathbb I}$), while the spectrum of an operator does not refer to a
representation: For any algebra element $\hat{a}\in{\cal A}$, it can be
defined as the set of all $\lambda\in{\mathbb C}$ such that
$\hat{a}-\lambda{\mathbb I}$ does not have an inverse in ${\cal A}$. The
main purpose of this article is to show that it is not only possible to define
the spectrum directly for an algebra, but also to compute it without using a
specific representation.

While this statement may seem formal, there are several useful implications
for physical considerations. In particular, (i) the algebraic derivation works
for all possible representations of the algebra, (ii) it applies equally to
pure states and mixed states, and (iii) it is available in systems of
non-associative quantum mechanics that cannot be represented on a Hilbert
space \cite{Malcev,JackiwMon,Jackiw}. The latter arena has recently led to a
new upper bound on the magnetic charge of elementary particles \cite{WeakMono}
and is therefore physically meaningful. Here, we demonstrate the new method
used in the latter result for standard associative systems, in which we
rederive known spectra but find new identities for moments of eigenstates that
can be interpreted as saturation conditions of higher-order uncertainty
relations. This result helps to demonstrate a relationship between excited
states and generalized coherent states.

Our starting point is the algebraic definition of a state as a (normalized)
positive linear functional on the $*$-algebra ${\cal A}$ of observables, that
is a linear map $\langle\cdot\rangle\colon{\cal A}\to{\mathbb C}$ with
$\langle\hat{a}^*\hat{a}\rangle\geq 0$ for all $\hat{a}\in{\cal A}$ (and
$\langle{\mathbb I}\rangle=1$). Physically, the positivity condition implies
not only that fluctuations
$\langle\hat{a}^2\rangle-\langle\hat{a}\rangle^2\geq 0$ of self-adjoint
algebra elements are positive, but also, and slightly less obviously, that
observations are subject to uncertainty relations; see for instance
\cite{LocalQuant}: Any positive state obeys the Cauchy--Schwarz inequality
\begin{equation}
 \langle\hat{a}^*\hat{a}\rangle \langle\hat{b}^*\hat{b}\rangle\geq
|\langle\hat{a}^*\hat{b}\rangle|^2
\end{equation}
from which uncertainty relations can be derived by making suitable choices for
$\hat{a}$ and $\hat{b}$.

The $*$-relation on ${\cal A}$ may be abstractly defined, or given by the
adjoint if ${\cal A}$ is represented on a Hilbert space. For basic generators
$\hat{x}_i$ of ${\cal A}$, such as positions and momenta, one can parameterize
the state by its basic expectation values $\langle\hat{x}_i\rangle$ and
central moments
\begin{equation}
 \Delta(x_1^{a_1}\cdots x_n^{a_n})= \langle
 (\hat{x}_1-\langle\hat{x}_1\rangle)^{a_1}\cdots
 (\hat{x}_n-\langle\hat{x}_n\rangle)^{a_n}\rangle_{\rm Weyl}
\end{equation}
using completely symmetric (or Weyl) ordering. Coupled equations of motion for
basic expectation values and moments follow from an extension of Ehrenfest's
theorem. For instance, for canonical $(x_i)=(q,p)$ with
$[\hat{q},\hat{p}]=i\hbar {\mathbb I}$, in addition to
\begin{equation}
\frac{{\rm d}\langle\hat{q}\rangle}{{\rm
    d}t}=\frac{\langle[\hat{q},\hat{H}]\rangle}{i\hbar} \quad,\quad 
\frac{{\rm d}\langle\hat{p}\rangle}{{\rm
    d}t}=\frac{\langle[\hat{p},\hat{H}]\rangle}{i\hbar} 
\end{equation}
we have
\begin{equation}
 \frac{{\rm d}\Delta(q^2)}{{\rm d}t} =  \frac{{\rm
     d}(\langle\hat{q}^2\rangle- \langle\hat{q}\rangle^2)}{{\rm d}t}
=
 \frac{\langle[\hat{q}^2,\hat{H}]\rangle}{i\hbar}- 2
 \langle\hat{q}\rangle\frac{{\rm d}\langle\hat{q}\rangle}{{\rm d}t}
\end{equation}
for the position variance $\Delta(q^2)=(\Delta q)^2$. Depending on the
Hamiltonian, the right-hand sides can be expanded in moments and usually
involve an asymptotic series of terms (unless the Hamiltonian is quadratic in
basic operators).

This formulation is especially useful for canonical effective theories
\cite{EffAc} and semiclassical expansions because the condition
$\Delta(x_1^{a_1}\cdots x_n^{a_n})=O(\hbar^{(a_1+\cdots+a_n)/2})$ provides a
general definition of semiclassical (but possibly non-Gaussian) states and
allows tractable approximations of the equations of motion order by order in
$\hbar$. In the present paper, as another new conceptual insight, we show that
interesting properties that can be obtained in this way are not restricted to
semiclassical ones: Harmonic and perturbative eigenvalues can be derived as
well, together with relationships between their moments.

Uncertainty relations play a crucial role in this context, as can be seen by
the simple example of the ground state of the harmonic oscillator with
Hamiltonian
\begin{equation}
 \hat{H}=\frac{1}{2m}\hat{p}^2+\frac{1}{2}m\omega^2\hat{q}^2\,.
\end{equation}
Using moments, the ground-state energy can be derived from two conditions,
namely that (i) the moments be time independent for a stationary state, and
(ii) the standard uncertainty relation be saturated. Indeed, in this case the
second-order moments obey a closed set of evolution equations
\begin{eqnarray}
\frac{{\rm d}\Delta(q^2)}{{\rm d}t} &=&  2\frac{\Delta(qp)}{m} \label{dDq}\\
\frac{{\rm d}\Delta(qp)}{{\rm d}t}&=&
\frac{1}{m} \Delta(p^2)-
m\omega^2 \Delta(q^2)\label{dDxp}\\ 
\frac{{\rm d} \Delta(p^2)}{{\rm d}t}&=& -2m\omega^2
\Delta(qp)\,. \label{dDp}
\end{eqnarray}
Condition (i) implies $\Delta(qp)=0$ and
$\Delta(p^2)=m^2\omega^2\Delta(q^2)$. Condition (ii) then determines
$\Delta(q^2)=\hbar/(2m\omega)$ and
$\Delta(p^2)=\frac{1}{2}m\omega\hbar$. Therefore, the energy expectation value
in such a state (with $\langle\hat{q}\rangle=0=\langle\hat{p}\rangle$ by
condition (i)), 
\begin{equation} \label{Hexp}
 \langle\hat{H}\rangle=\frac{1}{2m}
 \Delta(p^2)+\frac{1}{2}m\omega^2\Delta(q^2)= \frac{1}{2}\hbar\omega\,,
\end{equation}
agrees with the ground-state energy. It is not necessary to compute the full
ground-state wave function in order to find the energy.  However, the question
of how to compute the energy eigenvalues of excited states using moments is
more difficult: Their eigenstates are not Gaussian and therefore do
not saturate the standard uncertainty relation.

For the ground state of the harmonic oscillator, the condition that
Heisenberg's uncertainty relation be saturated can be replaced by a lesson
from the variational principle. The expectation value of the Hamiltonian is
minimized in the ground state. Since (\ref{Hexp}) is linear in second-order
moments, which take values in a region bounded by the uncertainty relation,
the expectation value is minimized at the boundary allowed by this
relation. Saturation therefore need not be assumed but can be derived from a
fundamental principle. But again, for excited states such a derivation based
on moments seems to be more complicated because one would somehow have to
restrict the moments to belong to a wave function orthogonal to the ground
state and all lower-excited states. However, orthogonality relations are not
available for states at the algebraic level. Our procedure will instead lead
to a selection of higher-order uncertainty relations which, regarding energy
eigenstates, split the state space into subsets much like the usual
orthogonality conditions do for wave functions.

For some time and in a slightly different context, moments have been known to
be useful for numerical approximations of eigenvalues of excited states
\cite{TransMom,MomentRecur,MomentMethod,MomentMethodGen}. (See also
\cite{ClassMoments,MomentsQuartic} for recent work.) Here, we use some of the
same relations between moments of eigenstates, but in a different way. As a
result, our constructions have a more fundamental flavor because they can
serve as new definitions of eigenvalues and eigenstates in the algebraic
perspective, even while they do provide new computational schemes as well. We
are aware of at least two examples for settings in which our constructions may
be useful: In canonical quantum gravity, the problem of time
\cite{KucharTime,Isham:Time,AndersonTime} often makes explicit constructions
of physical Hilbert spaces and wave functions untractable, while moment
methods have been shown to present certain computational advantages
\cite{EffTime,EffTimeLong,EffTimeCosmo,TwoTimes}.  And in non-associative
quantum mechanics, which plays a role in models with magnetic monopoles
\cite{MagneticCharge} or of certain flux compactifications in string theory
\cite{DualDoubled,NonAssGrav,NonGeoNonAss,TwistedNonAss,NonAssDef}, operators
on wave functions (and therefore the usual definition of eigenvalues) are in
general unavailable \cite{NonGeoNonAss,MSS1,BakasLuest,MSS2,MSS3}, but moments
may still be used \cite{NonAss,NonAssEffPot,WeakMono}.

The main new result we will be able to uncover here for associative systems is
a saturation property for any harmonic-oscillator eigenstate. (For a detailed
non-associative example, see \cite{HydroSpec}.) As part of our procedure, we
impose a set of inequality constraints involving the moments, so as to ensure
that they belong to an actual state (a {\em positive} linear
functional). These constraints include the standard uncertainty principle as
well as a series of inequalities involving higher moments. Upon imposing these
conditions, we find that some of them are not only satisfied but also
saturated by a harmonic-oscillator eigenstate. This feature is reminicient of
the saturation of Heisenberg's uncertainty relation by the ground state. As a
related result, we show that excited states of the harmonic oscillator are
(limits of) generalized coherent states as defined by Titulaer and Glauber
\cite{GenCoh}. In an extension to anharmonic oscillators, we confirm that such
saturation properties continue to hold order by order in perturbation theory
by the anharmonicity. Alternatively, eigenvalues can be derived from
convergence conditions for certain recurrence relations derived from
positivity and boundedness conditions of expectation values.

\section{Eigenvalues in a fermionic system}

As a warm-up, we compute eigenvalues in a fermionic system which has a
finite-dimensional Hilbert space in its standard representation, but we only
make use of the Grassmann algebra. The single degree of freedom $\xi$
included in this system is subject to anticommutation relations
\begin{equation}
 [\hat{\xi}^{\dagger},\hat{\xi}]_+=\hbar\quad,\quad
 [\hat{\xi},\hat{\xi}]_+=0=[\hat{\xi}^{\dagger},\hat{\xi}^{\dagger}]_+ \,.
\end{equation}
It generates a 4-dimensional unital $*$-algebra with vector-space basis given
by ${\mathbb I}$, $\hat{\xi}$, $\hat{\xi}^{\dagger}$ and
$\hat{\xi^{\dagger}}\hat{\xi}$. As a Hamiltonian, we choose
\begin{equation}
 \hat{H}=\frac{1}{2}\omega(\hat{\xi}^{\dagger}\hat{\xi}-
\hat{\xi}\hat{\xi}^{\dagger}) = \omega
\hat{\xi}^{\dagger}\hat{\xi}-\frac{1}{2}\hbar\omega {\mathbb I} = \omega
\hat{\xi}\hat{\xi}^{\dagger}+ \frac{1}{2}\hbar\omega {\mathbb I}\,.
\end{equation}

\subsection{Hilbert-space representation}

For comparison, we briefly summarize the standard representation on a
2-dimensional Hilbert space.  Commutators of $\hat{\xi}$ and
$\hat{\xi}^{\dagger}$ with $\hat{H}$ show that we can use the former as ladder
operators: we have $[\hat{\xi},\hat{H}]= \hbar\omega\hat{\xi}$. We define
$|-\rangle$ such that $\hat{\xi}|-\rangle=0$, and $|+\rangle$ as
$\hat{\xi}^{\dagger}|-\rangle=|+\rangle$. These two states are the only
independent ones since $\hat{\xi}^{\dagger}|+\rangle=(\hat{\xi}^{\dagger})^2
|-\rangle=0$. The eigenstates of $\hat{H}$ are then given by $|\pm\rangle$
with eigenvalues 
\begin{equation}
 E_{\pm}=\pm\frac{1}{2}\hbar\omega\,.
\end{equation}
The action of the ladder operators, $\hat{\xi}|+\rangle=
\sqrt{\hbar}|-\rangle$ and $\hat{\xi}^{\dagger}|-\rangle= \sqrt{\hbar}
|+\rangle$, follows from normalization of $|\pm\rangle$ and
\begin{eqnarray}
 ||\hat{\xi}|+\rangle||^2&=&\langle\hat{\xi}^{\dagger}\hat{\xi}\rangle_+=
\frac{1}{\omega}\left(E_++\frac{1}{2}\hbar\omega\right)=\hbar\\ 
||\hat{\xi}|-\rangle||^2&=& \langle\hat{\xi}\hat{\xi}^{\dagger}\rangle_-=
 \frac{1}{\omega} \left(-E_--\frac{1}{2}\hbar\omega\right)=\hbar\,.
\end{eqnarray}
A general state can be written as 
\begin{equation} \label{GeneralState}
 |r,s\rangle = \cos r|-\rangle+ e^{is} \sin r  |+\rangle\,,
\end{equation}
parameterizing all normalized states up to a phase. Expectation values in
these states are given by
\begin{eqnarray}
 \langle\hat{\xi}\rangle(r,s) &=& \frac{1}{2}\sqrt{\hbar} \sin(2r) e^{is} =
 \langle\hat{\xi}^{\dagger}\rangle(r,s)^*\\
\langle\hat{\xi}^{\dagger}\hat{\xi}\rangle(r,s) &=& \hbar \sin^2r\\
\langle\hat{\xi}\hat{\xi}^{\dagger}\rangle(r,s) &=& \hbar\cos^2r\,.
\end{eqnarray}

States are subject to uncertainty relations, which will play a major role in
our new method. Define $u=\Delta\hat{\xi} v$ and
$w=\Delta\hat{\xi}^{\dagger}v$ for some state $v$, where
$\Delta\hat{\xi}=\hat{\xi}- \langle\hat{\xi}\rangle_v$ with
$\langle\hat{\xi}\rangle_v=\langle v|\hat{\xi}v\rangle$, and compute
\begin{eqnarray}
 \langle u|u\rangle &=&
 \langle\Delta\hat{\xi}^{\dagger}\Delta\hat{\xi}\rangle =
 \Delta(\bar{\xi}\xi)+\frac{1}{2}\hbar\\
 \langle w|w\rangle &=&
 \langle\Delta\hat{\xi}\Delta\hat{\xi}^{\dagger}\rangle =
 -\Delta(\bar{\xi}\xi)+\frac{1}{2}\hbar\\
\langle u|w\rangle &=&
\langle\Delta\hat{\xi}^{\dagger}\Delta\hat{\xi}^{\dagger}\rangle =0
\end{eqnarray}
with the (graded) covariance
\begin{equation}
 \Delta(\bar{\xi}\xi)= \frac{1}{2}\left(\langle
 \hat{\xi}\hat{\xi}^{\dagger}-
 \hat{\xi}^{\dagger}\hat{\xi}\rangle- \langle\hat{\xi}\rangle
 \langle\hat{\xi}\rangle^*-\langle\hat{\xi}\rangle^*
 \langle\hat{\xi}\rangle\right)\,. 
\end{equation}
The Cauchy--Schwarz inequality implies
\begin{equation}
 0=|\langle u|w\rangle|^2\leq \langle u|u\rangle \langle w|w\rangle =
 -\Delta(\bar{\xi}\xi)^2+\frac{1}{4}\hbar^2
\end{equation}
and therefore
\begin{equation}
 |\Delta(\bar{\xi}\xi)|\leq \frac{1}{2}\hbar\,.
\end{equation}
Both eigenstates of $\hat{H}$ saturate this inequality.

\subsection{Algebra}

Let us now proceed algebraically. We introduce a phase-space version of the
fermion system by defining two complex numbers, $\xi=\langle\hat{\xi}\rangle$
and $\xi^*=\langle\hat{\xi}^{\dagger}\rangle$. The definition of a bracket
\begin{equation}
 \{\langle\hat{A}\rangle,\langle\hat{B}\rangle\}_+=
 \frac{\langle[\hat{A},\hat{B}]_+\rangle}{i\hbar}
\end{equation}
implies standard relations with anti-Poisson brackets
\begin{equation}
 \{\xi^*,\xi\}_+=-i\quad,\quad \{\xi,\xi\}_+=0=\{\xi^*,\xi^*\}_+
\end{equation}
for basic expectation values. The bracket can be extended to an anti-Poisson
bracket on moments of $\hat{\xi}$ and $\hat{\xi}^{\dagger}$ by using the
Leibniz rule.  

The basic expectation values also anticommute with $\hat{\xi}$ and
$\hat{\xi}^{\dagger}$ in products as they appear in moments. (This condition
is required for consistency with equations such as
$\langle\hat{\xi}\xi^*\rangle=\xi\xi^*$.)  There is only one non-zero moment:
\begin{equation}
 \Delta(\bar{\xi}\xi)= \frac{1}{2}
 \langle\Delta\hat{\xi}^{\dagger}\Delta\hat{\xi}-
 \Delta\hat{\xi}\Delta\hat{\xi}^{\dagger}\rangle 
 = \langle\Delta\hat{\xi}^{\dagger}\Delta\hat{\xi}\rangle-\frac{1}{2}\hbar =
 -\langle\Delta\hat{\xi}\Delta\hat{\xi}^{\dagger}\rangle+ \frac{1}{2}\hbar\,,
\end{equation}
using $\Delta\hat{\xi}:=\hat{\xi}-\xi$ and
$[\Delta\hat{\xi}^{\dagger},\Delta\hat{\xi}]_+=\hbar$.
The dynamics now follows from the usual derivation given by a commutator with
the Hamiltonian: 
\begin{equation}
 \dot{\xi} = \frac{\langle[\hat{\xi},\hat{H}]\rangle}{i\hbar} = -i\omega\xi
\end{equation}
implies $\xi(t)=\xi_0\exp(-i\omega t)$, or $r(t)=r_0$, $s(t)=s_0-\omega t$ in
the parameterization of (\ref{GeneralState}). Also,
$\Delta(\bar{\xi}\xi)(t)=\Delta(\bar{\xi}\xi)(0)$ because
$\Delta(\bar{\xi}\xi)=\omega^{-1}\hat{H}-|\xi|^2$ depends only on $\hat{H}$
and constants.

Assume now that we have an eigenstate of $\hat{H}$ with eigenvalue
$\lambda$. In this state,
\begin{eqnarray}
 0&=&\langle \hat{H}-\lambda {\mathbb I}\rangle = \omega
 \langle\hat{\xi}^{\dagger}\hat{\xi}\rangle- \frac{1}{2}\hbar\omega-\lambda
 = -\omega
 \langle\hat{\xi}\hat{\xi}^{\dagger}\rangle+\frac{1}{2}\hbar\omega-\lambda 
\label{H}\\
0&=& \langle\hat{\xi}(\hat{H}-\lambda {\mathbb I})\rangle =
\left(\frac{1}{2}\hbar\omega-\lambda\right)\xi \label{psi}\\
0&=& \langle\hat{\xi}^{\dagger}(\hat{H}-\lambda {\mathbb I})\rangle =
-\left(\frac{1}{2}\hbar\omega+\lambda\right) \xi^* \label{psid}\\
0&=& \langle\hat{\xi}^{\dagger}\hat{\xi}(\hat{H}-\lambda {\mathbb I})\rangle =
\left(\frac{1}{2}\hbar\omega-\lambda\right)
\langle\hat{\xi}^{\dagger}\hat{\xi}\rangle =
\frac{\frac{1}{4}\hbar^2\omega^2-\lambda^2}{\omega} \label{psidpsi}\\
0&=& \langle\hat{\xi}\hat{\xi}^{\dagger}(\hat{H}-\lambda {\mathbb I})\rangle =
-\left(\frac{1}{2} \hbar\omega+\lambda\right)
\langle\hat{\xi}\hat{\xi}^{\dagger}\rangle =
-\frac{\frac{1}{4}\hbar^2\omega^2-\lambda^2}{\omega} \label{psipsid}
\end{eqnarray}
using the first equation in the last step of (\ref{psidpsi}) and
(\ref{psipsid}). The last equation implies $\lambda_{\pm}=\pm
\frac{1}{2}\hbar\omega$. For $\lambda_-=-\frac{1}{2}\hbar\omega$, (\ref{psi})
implies $\xi=0$ and (\ref{psidpsi}) implies
$\langle\hat{\xi}^{\dagger}\hat{\xi}\rangle=0$, so that
$\langle\hat{\xi}\hat{\xi}^{\dagger}\rangle=\hbar$ from (\ref{H}). For
$\lambda_+=\frac{1}{2}\hbar\omega$, (\ref{psid}) implies $\xi^*=0$ and
(\ref{psipsid}) implies $\langle\hat{\xi}\hat{\xi}^{\dagger}\rangle=0$, so
that $\langle\hat{\xi}^{\dagger}\hat{\xi}\rangle=\hbar$ from (\ref{H}).

In this example, we have managed to compute all eigenvalues of the Hamiltonian
using only the (anti-)commutator relationships. If we try the standard method
of ladder operators in a system with an infinite-dimensional Hilbert space, it
is well known that we need normalizability conditions in order to derive
discrete eigenvalues. These conditions are available only for wave functions
in the Hilbert space but do not have an analog in the algebra of
observables. We now show that the new methods of using moments and uncertainty
relations can produce the correct discrete spectra without normalizability
condition even in systems with an infinite-dimensional Hilbert space.

\section{Eigenvalues from Moments}
\label{sec:eigs_and_mnt}

In this section, we rederive the well-known expressions for energy eigenvalues
of the harmonic oscillator based solely on relations between moments of an
eigenstate. For now, in order to highlight the mathematical features, we use
units such that the physical constants $m$, $\omega$ and $\hbar$ are equal to
one. Since the stationarity condition for energy eigenstates directly implies
that $\langle\hat{q}\rangle=0$ and $\langle\hat{p}\rangle=0$, we will work
directly with bare moments and zero basic expectation values.

\subsection{Notation}

We begin by introducing some notation. Define 
\begin{equation}
\hat{T}_{m,n} := (\hat{q}^m \hat{p}^n)_{\rm Weyl}
\end{equation}
where $\hat{q}$ and $\hat{p}$ are the canonical position and momentum
operators, $m$ and $n$ are non-negative integers, and the subscript indicates,
as before, that the product is taken in completely symmetric ordering. Note
that through the commutation relation $[\hat{q},\hat{p}] = i\hbar$, products
of the form $\hat{T}_{m,n} \hat{T}_{m',n'}$ can always be rewritten as sums
over individual $\hat{T}_{m'',n''}$ of order $m+n+m'+n'$ or less. See
\cite{OpDiff} for an explicit statement of the relevant reordering identity.

Given a particular state, we define the bare moments (about the origin) as:
\begin{equation}
T_{m,n} := \langle \hat{T}_{m,n}\rangle.
\end{equation}
The collection of all such moments for a given state provides a complete
description of the state in the sense that given the moments, it is possible
(in principle) to reconstruct the wave function. However, the moments are not
completely free. They must satisfy certain inequalities, such as Heisenberg's
uncertainty relation, as well as a number of other constraints involving
higher moments. A necessary and sufficient condition for a collection of
moments $\{T_{m,n}\}$ to correspond to a genuine quantum state has been given
in \cite{PositiveWigner}. More recently, a similar result has been developed
from a different perspective in \cite{MomentsUncert}, providing a generalized
uncertainty principle that imposes inequality constraints on higher
moments. These results are key for our further constructions.

Consider the column vector, $\hat{\bm{\xi}}_J$, consisting of all operators
$\hat{T}_{m,n}$ up to order $m+n=2J$, where $J$ is an integer or
half-integer. The generalized uncertainty principle states that the
$(2J+1)(2J+1)$ dimensional square matrix $M_J =
\langle\hat{\bm{\xi}}_J{\hat{\bm{\xi}}_J}^\dagger \rangle$ is positive
semi-definite,
\begin{equation}\label{uncertainty_princip}
M_J = \langle\hat{\bm{\xi}}_J{\hat{\bm{\xi}}_J}^\dagger \rangle \ge 0
\end{equation}
where the expectation value is taken element by element. Prior to taking the
expectation value, the matrix elements are products of the form $\hat{T}_{m,n}
\hat{T}_{m',n'}$. As mentioned above, these products can be rewritten as
linear combinations of individual $T_{m'',n''}$. The elements of $M_J$ are
thus functions of the moments. Since $M_J \ge 0$ implies non-negativity of its
principal minors, the generalized uncertainty principle yields a set of
inequalities involving the moments.

As discussed in \cite{MomentsWignerNonGauss}, it is useful to bring this
matrix to block diagonal form
\begin{equation} \label{block_uncertainty}
M_J \to
\left(\begin{array}{cccc}
A_0 	& 			&			&\\
 		& A_1		&		 	&\\
 		& 			&\ddots 	&\\
		&			&			& A_{2J}
\end{array}\right)
\end{equation}
where $A_n$ is an $n+1$ by $n+1$ matrix that contains moments up to order
$2n$. This can be achieved by repeatedly applying the following identity
\begin{equation} \label{block_identity}
L
\left(\begin{array}{cc}
A & C^\dagger\\
C & B
\end{array}\right)
L^\dagger 
= 
\left(\begin{array}{cc}
A & 0 \\
0 & B-C A^{-1}C^\dagger
\end{array}\right)
\end{equation}
 to $M_J$, where
\begin{equation}
L = 
\left(\begin{array}{cc}
1 & 0\\
-CA^{-1} & 1
\end{array}\right)\,.
\end{equation}
This identity holds whenever the matrix on the left-hand side of
Eq.~(\ref{block_identity}) is Hermitian. We then have that $M_J \ge 0$ if and
only if $A_n \ge 0$ for all $ n\le 2J$. The generalized uncertainty principle
may thus be rephrased as
\begin{equation}
A_n \ge 0 \hspace{3pt} \textrm{ for all } n \ge 0.
\end{equation}

If the state under consideration is known to be an eigenstate of a
Hamiltonian, $\hat{H}$, then we can obtain an additional set of
constraints. For all $m,n \ge 0$ we have
\begin{equation} \label{h_constraint}
\langle\hat{T}_{m,n}(\hat{H} - \lambda {\mathbb I}) \rangle_{\lambda} = 0
\end{equation}
where $\lambda$ is the eigenvalue of the state $\langle\cdot\rangle_{\lambda}$
under consideration. In order to rewrite this set of equations as a collection
of constraints on the moments, we express $\hat{H}$ in terms of the
$\hat{T}_{m,n}$ and reorder the product $\hat{T}_{m,n}\hat{H}$ into a sum over
individual $\hat{T}_{m',n'}$. Equation~(\ref{h_constraint}) then implies
recurrence relations for $T_{m,n}$ which depend on the system under
consideration.

\subsection{Application to the harmonic oscillator}
\label{sec:sho}

We now show how the considerations outlined above can be used to find the
eigenvalues of the harmonic-oscillator Hamiltonian. The idea is to use
(\ref{h_constraint}) to solve for the moments in terms of the eigenvalue
$\lambda$ and then apply (\ref{uncertainty_princip}) to obtain information
concerning the allowed values of $\lambda$ (as yet unspecified). This
combination is the basis of our new method.

\subsubsection{Recurrence relations}

For the sake of mathematical clarity, we use the Hamiltonian $\hat{H} =
(\hat{p}^2 + \hat{q}^2)/2$. The usual parameters given by the mass $m$ and
frequency $\omega$ can be reintroduced by a suitable canonical transformation
of $q$, $p$ if we also understand $\hat{H}$ as the energy divided by
$\omega$. Our $q$ and $p$ then both have units of $\sqrt{\hbar}$, such that
$T_{m,n}$ has units of $\hbar^{(m+n)/2}$.  Imposing (\ref{h_constraint})
results in the following relations between the moments
\begin{eqnarray}
T_{m+2,n} + T_{m,n+2} &=& 2\lambda T_{m,n}+\frac{n(n-1)}{4}\hbar^2
T_{m,n-2} + \frac{m(m-1)}{4}\hbar^2 T_{m-2,n}  \label{T1} \\
nT_{m+1,n-1} &=& mT_{m-1,n+1}\label{T2}
\end{eqnarray}
which hold for all $m,n\ge0$. Two constraints are obtained because
(\ref{h_constraint}) --- defined without symmetric ordering of the product
$\hat{T}_{m,n}\hat{H}$ --- has both real and imaginary parts. From (\ref{T2}),
starting with $m=0$ or $n=0$, we find that the moments are zero unless both
$m$ and $n$ are even. For even and non-zero $m=2j$ and $n=2k$, we then define
$S_{j,k}$ such that
\begin{equation}
 T_{2j,2k}=\frac{(2j)!(2k)!}{j!k!} S_{j,k}\,.
\end{equation}
For these coefficients,  (\ref{T2}) implies the simple relation
\begin{equation}
 S_{j+1,k}=S_{j,k+1}\,,
\end{equation}
which in turn implies that $S_{j,k}$ depends only on $j+k$. There are,
therefore, dimensionless coefficients $b_j$ depending only on a single
integer, such that
\begin{equation}
 T_{2j,2k}= \frac{(2j)!(2k)!}{j!k!}\hbar^{j+k} b_{j+k}\,.
\end{equation}
For convenience, it is useful to define a second set of coefficients, $a_j$,
such that 
\begin{equation}
 b_{j+k}=\frac{(j+k)!}{(2j+2k)!} a_{j+k}\,,
\end{equation}
or
\begin{equation} \label{m1}
T_{2j,2k}= \frac{(2j)!(2k)!(j+k)!}{j!k!(2j+2k)!} \hbar^{j+k} a_{j+k}\,.
\end{equation}
For instance, 
\begin{equation} \label{T2j0}
 T_{2j,0}=\hbar^ja_j
\end{equation}
and
\begin{equation} \label{T2j2}
 T_{2j,2}= \hbar^{j+1}\frac{a_{j+1}}{2j+1}
\end{equation}
have more compact coefficients than the equivalent expressions in terms of
$b_j$.

As a consequence of (\ref{T1}), the remaining coefficients, $a_{\ell}$, are
subject to a difference equation in a single independent variable:
\begin{equation} \label{m2}
a_{\ell+1} = \frac{\lambda\hbar^{-1}(2\ell+1)}{\ell + 1}a_\ell +
\frac{(2\ell+1)(2\ell)(2\ell-1)}{8(\ell + 1)}a_{\ell-1}\,.
\end{equation}
Given the two initial values $a_0=1$ (as a consequence of normalization of the
state, $T_{0,0}=1$) and $a_1=\lambda/\hbar$ (as a consequence of
$2\hbar a_1=T_{2,0}+T_{0,2}=2\langle\hat{H}\rangle_{\lambda}=2\lambda$), (\ref{m2})
determines all orders of moments in terms of the parameter $\lambda$.  It is
clear from the recurrence and its initial values that $a_{\ell}$ is a
polynomial in $\lambda$ of degree $\ell$. It has only even terms for $\ell$
even, and only odd terms for $\ell$ odd.

In terms of $b_{\ell}$, the recurrence relation is slightly simpler,
\begin{equation}
 (\ell+1)b_{\ell+1}- \frac{\lambda}{2\hbar} b_{\ell}-
 \frac{1}{16}\ell b_{\ell-1}=0\,,
\end{equation}
and can be solved via the generating function $f(x)=\sum_{\ell=0}^{\infty}
b_{\ell}x^{\ell}$ subject to the differential equation
\begin{equation}
 \left(1-\frac{1}{16}x^2\right)f'(x)=
 \frac{1}{2}\left(\frac{\lambda}{\hbar}+\frac{1}{8}x\right)f(x)
\end{equation}
and initial conditions $f(0)=b_0=1$, $f'(0)=b_1=\frac{1}{2}\lambda$. The
solution,
\begin{equation}
 f(x)=\frac{\left(1+x/4\right)^{\lambda/\hbar-1/2}}
{\left(1-x/4\right)^{\lambda/\hbar+1/2}}\,, 
\end{equation}
has the Taylor expansion
\begin{equation}
 f(x)= \sum_{\ell=0}^{\infty} \left(\frac{-x}{4}\right)^{\ell}
 \frac{(\ell-\lambda/\hbar-1/2)!}{(-\lambda/\hbar-1/2)!\ell!}
 {}_2F_1(\lambda/\hbar+1/2,-\ell; 
 \lambda/\hbar+1/2-\ell;-1)
\end{equation}
and determines the $b_{\ell}$ in terms of hypergeometric functions.

\subsubsection{Positivity}

We now apply the generalized uncertainty principle (\ref{uncertainty_princip})
to these moments. Note that $M_J \ge 0$ implies that $M_J' \ge 0$, where
$M_J'$ is a matrix formed by deleting from $M_J$ any number of rows and
their corresponding columns. Equivalently, $M_J'$ may be defined as the matrix
formed by deleting entries from $\hat{\bm{\xi}}_J$ to form a new vector
$\hat{\bm{\xi}}'_J$ and then taking
\begin{equation} \label{MJp}
 M'_J = \langle\hat{\bm{\xi}}'_J\hat{\bm{\xi}}'^\dagger_J \rangle\,.
\end{equation} 
In particular, consider the matrix $M'_J$ formed by taking $\hat{\bm{\xi}}'_J$
to contain only operators of the form $\hbar^{-m/2}\hat{T}_{m,0}$ and
$\hbar^{-m/2}\hat{T}_{m-1,1}$.

For example, for $J=0$ we have $M_0'=1$, not implying any non-trivial
uncertainty relation. For $J=1/2$, we have
\begin{equation}
 M'_{1/2}= \left\langle\left(\begin{array}{ccc} 1 &
       \hat{q}/\sqrt{\hbar}&\hat{p}/\sqrt{\hbar} \\
       \hat{q}/\sqrt{\hbar}&\hat{q}^2/\hbar&\hat{q}\hat{p}/\hbar\\
       \hat{p}/\sqrt{\hbar}&\hat{p}\hat{q}/\hbar&\hat{p}^2/\hbar 
\end{array}\right)\right\rangle 
\end{equation}
where the expectation value is taken element by element. A suitable minor of
$M'_{1/2}$ being positive semidefinite,
\begin{equation}
 \det\left(\begin{array}{cc}
\langle\hat{q}^2\rangle&\langle\hat{q}\hat{p}\rangle\\
\langle\hat{p}\hat{q}\rangle&\langle\hat{p}^2\rangle \end{array}\right)=
T_{2,0}T_{0,2}- \left(T_{1,1}+\frac{1}{2}i\hbar\right)
\left(T_{1,1}-\frac{1}{2}i\hbar\right)= T_{2,0}T_{0,2}- T_{1,1}^2-
\frac{\hbar^2}{4}\geq 0\,,
\end{equation}
is equivalent to Heisenberg's uncertainty relation. Taking $J=1$ as another
example, we have
\begin{equation}
\hat{\bm{\xi}}'_1 = 
\left(\begin{array}{c}
1 \\ \hat{T}_{1,0}/\sqrt{\hbar} \\ \hat{T}_{0,1}/\sqrt{\hbar} \\
\hat{T}_{2,0}/\hbar \\ \hat{T}_{1,1}/\hbar    
\end{array}\right)
\end{equation}
which gives
\begin{equation}
M'_1 = \left\langle
\left(\begin{array}{ccccc}
1 & \hat{T}_{1,0}/\sqrt{\hbar} & \hat{T}_{0,1}/\sqrt{\hbar} &
\hat{T}_{2,0}/\hbar & \hat{T}_{1,1}/\hbar  \\
\hat{T}_{1,0}/\sqrt{\hbar} & \hat{T}_{1,0}\hat{T}_{1,0}/\hbar &\hat{T}_{1,0}
\hat{T}_{0,1}/\hbar 
&\hat{T}_{1,0} \hat{T}_{2,0}/\hbar^{3/2} & \hat{T}_{1,0}\hat{T}_{1,1}/\hbar^{3/2} \\ 
\hat{T}_{0,1}/\sqrt{\hbar} & \hat{T}_{0,1}\hat{T}_{1,0}/\hbar &
\hat{T}_{0,1}\hat{T}_{0,1}/\hbar  &
\hat{T}_{0,1}\hat{T}_{2,0}/\hbar^{3/2} & \hat{T}_{0,1}\hat{T}_{1,1}/\hbar^{3/2} \\ 
\hat{T}_{2,0}/\hbar & \hat{T}_{2,0}\hat{T}_{1,0}/\hbar^{3/2} &
\hat{T}_{2,0}\hat{T}_{0,1}/\hbar^{3/2}  &
\hat{T}_{2,0}\hat{T}_{2,0}/\hbar^2 & \hat{T}_{2,0} \hat{T}_{1,1}/\hbar^2 \\ 
\hat{T}_{1,1}/\hbar & \hat{T}_{1,1}\hat{T}_{1,0}/\hbar^{3/2} &
\hat{T}_{1,1}\hat{T}_{0,1}/\hbar^{3/2}  &
\hat{T}_{1,1}\hat{T}_{2,0}/\hbar^2 & \hat{T}_{1,1}\hat{T}_{1,1}/\hbar^2 
\end{array}\right)
\right\rangle
\end{equation}
where as before the expectation value is taken element by element. 

In order to derive the generic structure of $M_J'$, we use the relations
\begin{eqnarray}
 \hat{T}_{k,0}\hat{T}_{\ell,1}&=& \hat{T}_{k+\ell,1}- \frac{1}{2}i k \hbar
 \hat{T}_{k+\ell-1,0}\\
 \hat{T}_{k,1}\hat{T}_{\ell,1}&=& \hat{T}_{k+\ell,2}+ \frac{1}{2}i
 (\ell-k)\hbar\hat{T}_{k+\ell-1,1}+ \frac{1}{4}k\ell\hbar^2 \hat{T}_{k+\ell-2,0}
\end{eqnarray}
which follow from the general ordering equations given in \cite{OpDiff} (or
\cite{MomentsWignerNonGauss}). Relabelling $\hat{\bm\xi}_J'$ as
\begin{equation}
 \hat{\bm\xi}_n' = \hbar^{-n/4}\cdot \left\{\begin{array}{cl} \hat{T}_{n/2,0} &
     \mbox{if }n\mbox{ 
       even}\\ \hbar^{1/4}\hat{T}_{(n-3)/2,1} & \mbox{if }n\mbox{
       odd}\end{array}\right. 
\end{equation}
we have
\begin{eqnarray}
&&\hat{M}_{mn}'= \hat{\bm\xi}_m' \hat{\bm\xi}_n'{}^{\dagger}=
\hbar^{-(m+n)/4}\cdot\left\{\begin{array}{cl} \hat{T}_{(m+n)/2,0} & \mbox{if }m,n
    \mbox{ even}\\ 
\hbar^{1/4}\hat{T}_{(m-3)/2,1}\hat{T}_{n/2,0} & \mbox{if }m\mbox{ odd and }n \mbox{
  even}\\
\hbar^{1/4}\hat{T}_{m/2,0}\hat{T}_{(n-3)/2,1} & \mbox{if }m\mbox{ even and }n\mbox{
  odd}\\
\hbar^{1/2}\hat{T}_{(m-3)/2,1}\hat{T}_{(n-3)/2,1}& \mbox{if }m,n\mbox{
  odd}\end{array}\right.\\
&=& \hbar^{-(m+n)/4}\cdot\left\{\begin{array}{cl} \hat{T}_{(m+n)/2,0} & \mbox{if
    }m,n \mbox{ even}\\ 
\hbar^{1/4}\hat{T}_{(m+n-3)/2,1}+\frac{1}{4}in\hbar^{5/4}\hat{T}_{(m+n-5)/2,0}
& \mbox{if }m\mbox{ 
  odd and }n \mbox{ even}\\
\hbar^{1/4}\hat{T}_{(m+n-3)/2,1}-\frac{1}{4}im\hbar^{5/4}
\hat{T}_{(m+n-5)/2,0} & \mbox{if }m\mbox{ 
  even and }n\mbox{ odd}\\
\hbar^{1/2}\hat{T}_{(m+n-6)/2,2}+ \frac{n-m}{4}i\hbar^{3/2} \hat{T}_{(m+n-8)/2,1}&\\
\qquad\qquad  +
\frac{(m-3)(n-3)}{16}\hbar^{5/2} \hat{T}_{(m+n-10)/2,0}&  \mbox{if }m,n\mbox{
  odd}\end{array}\right. \nonumber
\end{eqnarray}
Taking expectation values and setting all $T_{m,n}=0$ unless $m$ and $n$ are
even, we obtain
\begin{equation}
M_{mn}'=\hbar^{-(m+n)/4}\cdot\left\{\begin{array}{cl} T_{(m+n)/2,0} & \mbox{if
    }m,n \mbox{ even}\\ 
\frac{1}{4}in\hbar^{5/4} T_{(m+n-5)/2,0} & \mbox{if }m\mbox{
  odd and }n \mbox{ even}\\
-\frac{1}{4}im\hbar^{5/4} T_{(m+n-5)/2,0} & \mbox{if }m\mbox{
  even and }n\mbox{ odd}\\
\hbar^{1/2}T_{(m+n-6)/2,2}+ 
\frac{(m-3)(n-3)}{16}\hbar^{5/2} T_{(m+n-10)/2,0}& \mbox{if }m,n\mbox{
  odd}\end{array}\right.
\end{equation}

Some components of $M_{mn}'$ are zero for certain values of $m$ and $n$, which
can be seen by refining the parameterization such that $m=4q+\alpha$ and
$n=4r+\beta$ with integer $q$ and $r$ and $0\leq\alpha,\beta\leq 3$. For fixed
$q$ and $r$, we obtain the $4\times 4$ block
\begin{eqnarray}
 &&\hbar^{q+r}M'_{4q+\alpha,4r+\beta}=\\
&& \left(\begin{array}{cccc} T_{2(q+r),0}&
     -iq\hbar T_{2(q+r-1),0}&0&0\\ ir\hbar T_{2(q+r-1),0} & T_{2(q+r-1),2}& 0&0\\ 
&+  (q-\frac{1}{2})(r-\frac{1}{2})\hbar^2 T_{2(q+r-2),0} &&\\
0&0&\hbar^{-1}T_{2(q+r+1),0}&
     -i(q+\frac{1}{2})T_{2(q+r),0}\\ 0&0& i(r+\frac{1}{2}) T_{2(q+r),0}&
     \hbar^{-1}T_{2(q+r),2}\\&&&+ 
     qr\hbar T_{2(q+r-1),0}\end{array}\right)\nonumber
\end{eqnarray}
where rows and columns are arranged according to the values of $\alpha$ and
$\beta$. (The full $4\times 4$-blocks appear in $M_J'$ only for $q\geq 1$ and
$r\geq 1$, while parts of these blocks make up the first three rows and
columns of $M_J'$.) Using (\ref{T2j0}) and (\ref{T2j2}), we obtain the blocks
\begin{eqnarray}
 &&\hbar^{q+r}M'_{4q+\alpha,4r+\beta}= \\
&&\left(\begin{array}{cccc} a_{q+r}&
     -iqa_{q+r-1}&0&0\\ ira_{q+r-1} &  \frac{1}{2(q+r)-1}a_{q+r}+
     (q-\frac{1}{2})(r-\frac{1}{2})a_{q+r-2}& 0&0\\ 0&0&a_{q+r+1}&
     -i(q+\frac{1}{2})a_{q+r}\\ 0&0& i(r+\frac{1}{2})a_{q+r}&
     \frac{a_{q+r+1}}{2(q+r)+1}+ 
     qra_{q+r-1}\end{array}\right)\nonumber
\end{eqnarray}

If $J=1$, for instance, we have the matrix
\begin{equation}
M_1'= \left(\begin{array}{ccccc}1&0&0&a_1&0\\ 0&a_1&\frac{1}{2}i&0&0\\
    0&-\frac{1}{2}i&a_1&0&0\\ a_1&0&0&a_2&i a_1\\ 0&0&0&-i
    a_1& \frac{1}{3}a_2+\frac{1}{4}\end{array}\right)\,.
\end{equation}
It is block-diagonalized by identifying $C^{\dagger}$ in
(\ref{block_identity}) with the vector $C_1^{\dagger}=(0,0,a_1,0)$:
\begin{equation}
L_1M_1'L_1^{\dagger}= \left(\begin{array}{ccccc}1&0&0&0&0\\
    0&a_1&\frac{1}{2}i&0&0\\ 
    0&-\frac{1}{2}i&a_1&0&0\\ 0&0&0&a_2-a_1^2&i a_1\\ 0&0&0&-i
    a_1& \frac{1}{3}a_2+\frac{1}{4}\end{array}\right)\,.
\end{equation}
Its determinant is equal to
\begin{equation} \label{det1}
 \det(L_1M_1'L_1^{\dagger})= \frac{1}{4}
 (\lambda/\hbar+1/2)^2(\lambda/\hbar-1/2)^2(\lambda/\hbar+3/2)
(\lambda/\hbar-3/2) 
\end{equation}
using the solution $a_2=\frac{3}{2}(\lambda^2/\hbar^2+1/4)$ of the recurrence
relation (\ref{m2}). 

\subsubsection{Eigenvalues}

For any $J$, we may block diagonalize $M_J'$ as in Equation
(\ref{block_uncertainty}), except that each $A_n'$ will be a $2\times2$ matrix
since we are working with the reduced matrix, $M_J'$. We then have
\begin{equation} \label{det_condition}
\det(A_n') \ge 0\,
\end{equation}
for all $n$. For a fixed $n$, this is a constraint involving moments up to
order $2n$. All of these moments can in turn be written in terms of $\lambda$
using (\ref{m1}) and (\ref{m2}). From explicit computations, we infer the
general result
\begin{equation}\label{det}
d_n=\det(A_n') = \frac{1}{4^{n-1}} \prod_{k=1}^n (\lambda/\hbar -
\alpha_k)(\lambda/\hbar + \alpha_k) 
\end{equation}
where $\alpha_k = (2k-1)/2$ are the odd half-integer multiples. (The
polynomial (\ref{det1}) is equal to $d_1d_2$.) Considered as a function of
$\lambda$, this expression has nodes at the $\alpha_k$ up to some maximum $k$
that depends on the particular value of $n$. Between nodes, the function is
non-zero, and it alternates in sign depending on the value of $n$. In
particular, because $d_{n+1}=\frac{1}{4}d_n(\lambda^2/\hbar^2-\alpha_k^2)$
implies ${\rm sgn}\,d_{n+1}=-{\rm sgn}\,d_n$ if $|\lambda|/\hbar<\alpha_n$,
sending $n \to n+1$ causes the sign to alternate. This behavior combined with
the non-negativity of $\det(A_n')$ implies that the only allowable values for
$\lambda$ occur at the nodes. We can exclude negative values of $\lambda$ by
appealing to the non-negativity of the first leading principal minor of $A_1'$
(which in this case is a $1\times1$ ``block'' consisting simply of $\lambda$),
which gives the constraint $\lambda \ge 0$. We thus have that the only
possible values for $\lambda$ are
\begin{equation} \label{lambda}
\lambda = \frac{1}{2}\hbar,\frac{3}{2}\hbar,\frac{5}{2}\hbar, \ldots
\end{equation}
in agreement with the well-known eigenvalues of the harmonic-oscillator
Hamiltonian (divided by $\omega$).

Since eigenvalues occur at the nodes of positivity conditions, all excited
states obey saturation conditions of higher-order uncertainty relations. We
will explore these relations further in Section~\ref{sec:saturation}, but
first give an alternative moment-based derivation of eigenvalues because we
have found it to be difficult to construct a general analytic proof of our
crucial equation (\ref{det}).

\subsection{Alternative derivation}

Given an energy eigenstate of the harmonic oscillator with eigenvalue
$\lambda$,  consider  the function
\begin{equation} \label{Ldef}
L_{\lambda}( \gamma ) = \left\langle\exp\left((1+\gamma) \hat{q}^2/\hbar
  \right) \right\rangle_{\lambda}\,. 
\end{equation}
For fixed $\lambda$, this function of $\gamma$ is well defined for $\gamma
\leq -1$ because $\exp\left((1+\gamma) \hat{q}^2/\hbar \right)$ is then an
algebra element that quantizes a bounded function, with $L_{\lambda}(-1) = 1$
by normalization and $\lim_{\gamma \to -\infty} L_{\lambda}(\gamma) = 0$. (Any
positive state is continuous.) Positivity of the state also implies that
$L_{\lambda}(\gamma)$ increases monotonically. We will show that these
properties, implied by boundedness and positivity, can replace the uncertainty
relations used in the preceding section in an algebraic derivation of
eigenvalues.

\subsubsection{Recurrence relations}

The moment expansion
\begin{eqnarray*}
  L_{\lambda}(\gamma) &=& \sum_{j=0}^{\infty}\hbar^{-j}
  \langle\hat{q}^{2j}\rangle_{\lambda} 
  \frac{(1+\gamma)^j}{j!} \\ 
  &=&  \sum_{j=0}^{\infty} a_{j} \frac{(1+\gamma)^j}{j!}
\end{eqnarray*} 
is readily obtained from the Taylor series of the exponential function,
followed by the identification
$\hbar^{-j}\langle\hat{q}^{2j}\rangle=\hbar^{-j}T_{2j,0}=a_j$ according to
(\ref{T2j0}).  Using the recursion relation (\ref{m2}) for the $a_j$ we obtain
the differential equation
\begin{eqnarray*}
3 L_{\lambda} + 3(9 + 9 \gamma + 4 \lambda/\hbar) L'_{\lambda} + 8(2+
\lambda/\hbar + 
\gamma (6 + 3 \gamma + \lambda/\hbar)) L''_{\lambda} + 4 \gamma
(1+\gamma)(2+\gamma) 
L'''_{\lambda} = 0 
\end{eqnarray*}
where primes indicate derivatives by $\gamma$.  Motivated by the behavior of
$L_{\lambda}(\gamma)$ as $\gamma\to-\infty$, we rewrite this function as
\begin{equation} \label{Lalpha}
 L_{\lambda}(\gamma)  
= \sum_{n=0}^{\infty} \alpha_{n,s} (-\gamma)^{-n - s}
\end{equation} 
where the constant $s$ takes into account a possible root-like pole at
$\gamma\to-\infty$. The $\alpha_{n,s}$ are then subject to the relation
\begin{eqnarray*}
  8 (n+s) (n+s - \lambda/\hbar) \alpha_{n,s} - (1+ 2n + 2 s) \biggl((3+6n+6s - 4
    \lambda/\hbar) 
    \alpha_{n+1,s} - (3+2n + 2 s)\alpha_{n+2,s}\biggr) = 0 \,.
\end{eqnarray*}
Inserting $n=-1$ and requiring that this sequence of numbers terminates before
$n=0$ in backwards recurrence implies $s=\frac{1}{2}$. With this knowledge we
can rewrite $L$ as
\begin{eqnarray} \label{LA} 
L_{\lambda}(\gamma) = \sum_{n=0}^{\infty}  A_{n} (-\gamma)^{-n - \frac{1}{2}}
\end{eqnarray}
where $A_{n}=\alpha_{n,1/2}$. The preceding recurrence relation
then turns into
\begin{eqnarray} \label{An}
(1+2 n) (1+ 2n  - 2 \lambda/\hbar) A_{n} - 2(1+ n) \biggl((3+3n - 2
\lambda/\hbar) A_{n+1} - 
(2+n) A_{n+2}\biggr) = 0 \,.
\end{eqnarray}

In the large-$n$ limit, equation~(\ref{An}) simplifies to $4 A_{n} - 6 A_{n+1}
+ 2 A_{n+2} = 0$. Therefore, for very large $n$, $A_{n} \approx c_{1} 
+ 2^{n} c_2$. If $c_1\not=0$ or $c_2\not=0$, this asymptotic behavior is
problematic as it would cause
\begin{eqnarray}
L_{\lambda}(\gamma) &\approx& \sum_{n= 0}^{M-1} A_{n}
  (-\gamma)^{-n-\frac12}  + \sum_{n= M}^{\infty} \left( c_{1}
  (-\gamma)^{-n-\frac12} + 2^nc_2 (-\gamma)^{-n-\frac12}\right)\nonumber\\ 
&=& \sum_{n= 0}^{M-1}  A_{n} (-\gamma)^{-n-\frac12}  -
(-\gamma)^{\frac12-M} \left( \frac{c_{1}}{1+\gamma}+\frac{
    2^{M}c_2}{2+\gamma} \right) 
\end{eqnarray}
to diverge on values of $\gamma$, $\gamma=-1$ and $\gamma=-2$, where it ought
to be between zero and one. 

Therefore, both $c_1$ and $c_2$ have to be strictly zero: after a certain $n$
all the $A_{n}$ should vanish. Let $N$ be the lowest integer such that $A_N =
0$. (Such an $N$ always exists because the normalization condition $L_\lambda
(-1)=1$ cannot be satisfied if all $A_{n}$ are zero.) We then obtain the
consistency equation
\begin{eqnarray*}
(2N - 1) (2N - 1 - 2 \lambda/\hbar) A_{N-1} = 0
\end{eqnarray*}
from inserting $n = N-1$ in (\ref{An}). By definition $A_{N-1}$ is
nonzero. Combined with the fact that $N$ is an integer greater than zero, we
find the familiar spectrum (\ref{lambda}).

\subsubsection{Coefficients}

Based on this result, the coefficients introduced in (\ref{LA}) seem to be
more tractable in the eigenvalue problem compared with our original
$a_j$. These sets are strictly related to each other, but not in a simple way.
Using Cauchy's formula to invert (\ref{LA}), we first write
\begin{eqnarray} 
A_n &=& \frac{1}{2 \pi i} \oint_{|z|=1} L_{\lambda}(z) z^{n -\frac12} \mathrm{d}z
 = \sum\limits_{j=0}^{\infty} \frac{a_{j}}{2 \pi j!} \int\limits_{0}^{2\pi}
 (1 + e^{i \theta})^j e^{i (n+ 1/2) \theta} \mathrm{d}\theta \nonumber\\ 
&=& \sum\limits_{j=0}^{\infty} \frac{(-1)^n a_{j}}{\pi j!}
B(-1;n+1/2,j+1)\label{Anseries} 
\end{eqnarray}
using also (\ref{Lalpha}), where $B$ is the incomplete beta function. 

In order to check convergence, we write $(1+e^{i\theta})^j=2e^{ij\theta/2}
\cos(\theta/2)^j$ and show that the second factor can be approximated as
$\cos(\theta/2)^j \approx \exp(-j \theta^2/8)$. It is straightforward to
confirm that these two expressions match to second order of a Taylor expansion
in $\theta$ around $\theta=0$. The local maxima of the difference of
$\cos(\theta/2)^j$ and $\exp(-j \theta^2/8)$ are at $\theta_{\rm max}$ such that
\begin{eqnarray*}
  0 &=& \partial_{\theta} \left(\cos(\theta/2)^j - \exp(-j \theta^2/8)
  \right)_{\theta = \theta_{\rm max}} \\ 
  &=& \frac{j}{4} \left(\theta_{\rm max} \exp(-j \theta_{\rm max}^2/8)  - 2
    \tan(\theta_{\rm max}/2) \cos(\theta_{\rm max}/2)^j\right)  
\end{eqnarray*}
such that
\[
 \cos(\theta_{\rm max}/2)^j= \frac{\theta_{\rm max}/2}{\tan(\theta_{\rm
     max}/2)} \exp(-j\theta_{\rm max}^2/2)\,.
\]
Therefore, the difference is bounded by
\begin{eqnarray*}
  \Delta_j &:=& \sup_{\theta \in [-\pi,\pi]}|\cos(\theta/2)^j - \exp(-j
  \theta^2/8)| = |\cos(\theta_{\rm max}/2)^j - \exp(-j
  \theta_{\rm max}^2/8)|\nonumber\\
&=&
  \left(1 - \frac{\theta_{max}/2}{\tan(\theta_{max}/2)}\right)
  \exp(-j \theta_{max}^2/8)  \,.
\end{eqnarray*}
This expression goes to zero for large $j$ because of the exponential factor,
unless $\theta_{\max}\to 0$ in which case the first factor in $\Delta_j$
approaches zero. We conclude that the difference of the two functions
$\cos(\theta/2)^j$ and $\exp(-j \theta^2/8)$ converges to zero in
$L^{\infty}[-\pi,\pi]$ when $j$ goes to infinity.

Approximating $(1+ e^{i \theta})^j \approx 2^j \exp(- j \theta^2/8+
\frac{i}{2} j \theta)$ in the incomplete beta function, we have
\begin{eqnarray*}
B(-1;n+1/2,j+1) &=& \frac{(-1)^n}{2} \int_{-\pi}^{\pi} (1 + e^{i
  \theta})^j e^{i (n+ 1/2) \theta} \mathrm{d}\theta \\ 
&\leq&  \frac{(-1)^n}{2} \int^{\infty}_{-\infty} 2^j \exp(- j \theta^2/8+
i j \theta/2) e^{i (n+ 1/2) \theta} \mathrm{d}\theta + 2^{j+1} \pi
\Delta_j\\ 
&=& 2\pi (-1)^n \frac{2^j}{j}  \exp\left(-\frac{(1+ j + 2 n)^2}{2 j}\right) +
2^{j+1} \pi \Delta_j \,.
\end{eqnarray*}
The first term goes to zero for fixed $n$ and large $j$. From the recursion
relation for the $a_j$, we then see that the series (\ref{Anseries}) for
$A_{n}$ has to converge as well, as the numerator grows at most
exponentially with $j$, while the denominator contains a $j!$.

Conversely, we have
\begin{eqnarray*}
a_{j}  &=& \left.\left(\frac{\mathrm{d}^j}{\mathrm{d} \gamma^j}
L_{\lambda}(\gamma)\right)\right\vert_{\gamma = -1} \\ 
&=& \sum_{n=0}^{\infty}  A_{n} \left.\left(\frac{\mathrm{d}^j}{\mathrm{d} \gamma^j}
  (-\gamma)^{-n - \frac12}\right) \right\vert_{\gamma = -1} \\ 
&=& (-1)^{j}\sum_{n=0}^{\infty} A_{n}  \left(-n-\frac12\right)^{(j)} \\
\end{eqnarray*}
where $x^{(n)}$ is the $n$th Pochhammer polynomial. As we have seen, only a
finite number of the $A_n$ are nonzero, and therefore this sum is clearly well
defined.

\subsubsection{Probability density}

The alternative method based on (\ref{Ldef}) allows a more direct derivation
of the probability density of eigenstates compared with reconstruction from
the moments of Section~\ref{sec:sho}.  

In order to reconstruct the probability density of the $N^{\rm th}$ energy
level, we first solve the recurrence relation for the coefficients $A_n$. Once
$N$ is fixed for a given eigenstate, we know that the $N^{\rm th}$
coefficient, $A_N$, is the highest non-zero one Its exact value will be fixed
later by normalization. Running through the recursion relation (\ref{An}) with
the known eigenvalue $\lambda = \hbar (N+\frac{1}{2})$, we can then work
backward, starting with $n=N-1$, until we reach the $0^{\rm th}$ coefficient
$A_0$ using (\ref{An}) for $n=0$. After that, the recurrence terminates
automatically: For $n=-1$ in (\ref{An}), we obtain $A_{-1}=0$ because of an
overal factor of $(1+n)$ in the second part of (\ref{An}), and for $n=-2$ we
obtain $A_{-2}=0$ because $A_{-1}$ is zero, as just shown, and there is a
factor of $(n+2)$ in front of the $A_{0}=A_{n+2}$ in this case. All
coefficients of orders less than $-2$ then vanish because the recurrence is of
second order.  As an example, we consider $N=4$ and find
\begin{eqnarray*}
A_3 &=& -\frac{12}{7} A_4 \\
A_2 &=& \frac{6}{5} A_4 \\
A_1 &=& -\frac{12}{35}A_4\\
A_0 &=& \frac{3}{5}A_4\,.
\end{eqnarray*}

The coefficients $A_n$ then determine the function $L_{\lambda}(\gamma)$, in
which we can impose normalization by requiring $L_{\lambda}(-1) =
\langle{\mathbb I}\rangle_{\lambda}=1$. Continuing with our example of $N=4$,
we find
\begin{equation}
L_{\lambda_{4}} =  \frac{35 + 60 \gamma + 42 \gamma^2 + 12 \gamma^3 + 3
  \gamma^4}{8(-\gamma)^{9/2}} \,.
\end{equation}
The probability density then requires an inversion of the integral that
defines the expectation value taken in $L_{\lambda}(\gamma)$.

In order to do so, we first note that the Hamiltonian commutes with the parity
operator, such that the probability density of any eigenstate has to be
even. We therefore write
\begin{equation}
 L_{\lambda}(\gamma) = 2 \int_{0}^{\infty} \exp\left(\frac{1+\gamma}{\hbar}
   x^2\right) P_{\lambda}(x)\mathrm{d}x
\end{equation}
in order to introduce the probability density $P_{\lambda}(x)$.  Subsituting
$u=x^2$ and $t =-\frac{1+\gamma}{\hbar}$, where all are expressions are
well-defined if ${\rm Re}(t)>0$, we obtain 
\begin{equation}
 L_{\lambda}(-1-\hbar t) = \int_{0}^{\infty} e^{-tu}
 \frac{P_{\lambda}(\sqrt{u})}{\sqrt{u}}\mathrm{d}u\,.
\end{equation}
The probability density is therefore obtained by an inverse Laplace transform,
for which we can use Mellin's inverse formula (with a suitable $\delta$):
\begin{eqnarray}
P_{\lambda}(x) &=& \frac{x}{2 \pi i} \lim_{T \to \infty} \int_{\delta- i
  T}^{\delta + i T} e^{t x^2} L_{\lambda}(-1-\hbar t) \mathrm{d}t \nonumber\\ 
&=& \sum_{n=0}^{N} \frac{x}{2 \pi i} \lim_{T \to \infty} \int_{\delta- i
  T}^{\delta + i T} e^{t x^2} A_{n} (1+\hbar t)^{-n -\frac12}\mathrm{d}t \nonumber\\ 
&=& \sum_{n=0}^{N}\frac{ A_{n} n! (2 x)^{2n}
  \exp(-x^2/\hbar)}{\sqrt{\pi}(2n)!\hbar^{n+\frac12}}\,.
\end{eqnarray} 
Proceeding again for our example of $N=4$, we have
\begin{eqnarray}
P_{\lambda_{4}}(x) &=&  \frac{\exp(-x^2/\hbar)}{\sqrt{\pi
    \hbar}}\left(\frac{3}{8} -\frac{12}{8} \frac{2 x^2}{\hbar} +\frac{42}{8}
  \frac{4 x^4}{3 \hbar^2}- \frac{60}{8} \frac{8 x^6}{15
    \hbar^3}+\frac{35}{8}\frac{16 x^8}{105\hbar^4}\right) \nonumber\\ 
&=& \frac{\exp(-x^2/\hbar)}{24\sqrt{\pi\hbar}} \left(3 - 12 \frac{x^2}{\hbar}
  + 4 \frac{x^4}{\hbar}\right)^2 \nonumber\\ 
&=& \frac{\exp(-x^2/\hbar)}{\sqrt{\pi\hbar} 2^4 4!}
H_{4}\left(\frac{x}{\sqrt{\hbar}}\right)^2 = |\psi_{4}(x)|^2\,.
\end{eqnarray}

The method introduced in the present subsection is more efficient than the
moment method, and perhaps more powerful because it provides a more direct
route to probability densities of eigenstates. However, the key definition
(\ref{Ldef}) of the function $L_{\lambda}(\gamma)$ was made with the benefit
of knowing that the operator $\exp((1+\gamma)\hat{q}^2/\hbar)$ should be
useful, based on the known form of wave function for harmonic-oscillator
eigenstates. While this alternative method is fully algebraic, just like the
moment method, it is not completely independent of standard derivations of
eigenstates. 

We note at this point that other algebraic derivations of eigenvalues and
eigenstates of the harmonic oscillator exist in the literature, such as
\cite{HarmOscAlgebraic}. However, they are based on ladder operators in
Hilbert space and therefore require representations of the algebra of
observables.

\section{Saturation of inequalities}
\label{sec:saturation}

An interesting result that emerges from the solutions in Section~\ref{sec:sho}
is a saturation property of the first $n$ eigenstates that obey $d_n=0$, and
therefore saturate the generalized uncertainty relation $\det(A_n') \ge 0$
given in (\ref{det}).  For $n=1$, this condition is just the well-known
statement that the harmonic-oscillator ground state saturates Heisenberg's
uncertainty relation. For each $n>1$, we have an inequality involving higher
moments that is saturated by the first $n$ eigenstates. (This saturation
property is different from the one found in \cite{HarmonicSaturate}. Moreover,
it sharpens a saturation property found in \cite{MomentsWignerNonGauss}, which
is true for all energy eigenstates of the harmonic oscillator.)  Motivated by
this finding, we return to the full generalized uncertainty principle and
analyze its behavior for the harmonic oscillator eigenstates, as well as
related properties.

\subsection{Principal minors and pure states}

As is evident from our derivations in the previous section, we need to make
use of only a submatrix of $M_J$, corresponding to moments in
$\hat{\bm{\xi}}_J'$ with at most one insertion of a momentum operator. (A
related computational fact is that $M_J$ has an eigenvalue zero with
degeneracy $D=J(2J-1)$.)  Computational experiments indicate that the
remaining conditions do not impose additional restrictions on the allowed
values of $\lambda$, which is consistent with the fact that (\ref{lambda}) is
the full set of harmonic-oscillator eigenvalues.

Still, for an application of the method without prior knowledge of the
spectrum, it would be of interest to understand these features in more
detail. In particular, it remains unclear to us how a suitable subset of
independent inequalities can be selected from the generalized uncertainty
principle that would be sufficient for determining all eigenstates of a given
Hamiltonian.

The observation that the matrices $M_J'$ suffice to find all relevant
conditions on eigenvalues can be interpreted as follows: For pure states, the
moments $T_{m,0}=\langle\hat{q}^m\rangle$ allow one to reconstruct the norm of
the wave function according to the Hamburger problem, while the additional
moments $T_{n,1}=\langle\hat{q}^n\hat{p}\rangle$ with a single momentum
operator can be used to determine the phase; see for instance
\cite{EffAc,EffPotRealize}. The other moments are therefore not independent
parameters if the state is known to be pure. (They would be independent for
mixed states.) The observation that $M_J'$ suffices to find all conditions on
eigenvalues, at least for the harmonic oscillator, can therefore be
interpreted as saying that mixed states cannot provide eigenstates in this
case.

\subsection{Saturation from ladder operators}

With hindsight, it is possible to obtain a saturation result for energy
eigenstates of the harmonic oscillator by means of the usual ladder operators,
\begin{equation}
 \hat{a}= \frac{1}{\sqrt{2\hbar}} (\hat{q}+i\hat{p}) \quad,\quad
\hat{a}^{\dagger}= \frac{1}{\sqrt{2\hbar}} (\hat{q}-i\hat{p})\,.
\end{equation}
(We still assume $m=1$ and $\omega = 1$.)  Let $\hat{a}$ be the lowering
operator and take
\begin{equation} \label{ladder_ops}
\hat{f} = \hat{a}^n + \hat{a}^\dagger{}^n\quad,\quad
\hat{g} = \hat{a}^n - \hat{a}^\dagger{}^n\,.
\end{equation}
If a state $|\psi\rangle$ is a linear combination of the first $n$
eigenstates of the harmonic oscillator, then $\hat{f}|\psi\rangle$ =
$-\hat{g}| \psi \rangle$, which implies
$\langle\hat{f}^{\dagger}\hat{f}\rangle \langle
\hat{g}^{\dagger}\hat{g}\rangle = \langle\hat{f}^{\dagger}\hat{g}\rangle
\langle \hat{g}^{\dagger}\hat{f}\rangle$. Thus, the Cauchy-Schwarz
inequality
\begin{equation} \label{cs_saturated}
\langle \hat{f}^\dagger \hat{f}\rangle \langle \hat{g}^\dagger \hat{g} \rangle
\ge |\langle \hat{f}^\dagger \hat{g} \rangle|^2 
\end{equation}
is saturated. Explicit expressions for given $n$ imply higher-order
uncertainty relations, which must then also be saturated by the first $n$
energy eigenstates of the harmonic oscillator.

The first three inequalities obtained in this way are as follows.  The $n^{\rm
  th}$ inequality is saturated by any linear combination of the first $n$
harmonic-oscillator eigenstates. For
$n=1$,
\begin{equation}
\left\langle\hat{q}^2 \right\rangle \left\langle \hat{p}^2 \right\rangle \ge
\hbar^2/4+\left\langle\hat{q}\hat{p} \right\rangle_{\rm Weyl} ^2 
\end{equation}
for $n=2$,
\begin{eqnarray}
&&\Bigg(\langle \hat{p}^4 \rangle + \langle \hat{q}^4\rangle - 2 \langle
\hat{p}^2 \hat{q}^2 \rangle_{\rm Weyl}  +
\hbar^2\Bigg)\Bigg(\langle\hat{p}^2\hat{q}^2 
\rangle_{\rm Weyl}+ \frac{\hbar^2}{4}\Bigg) \\ \nonumber 
&&\ge \hbar^2\Bigg(  \langle \hat{p}^2 \rangle + \langle \hat{q}^2
\rangle\Bigg)^2 + \Bigg(  \langle\hat{p}\hat{q}^3 \rangle_{\rm Weyl} -  \langle
\hat{p}^3\hat{q} \rangle_{\rm Weyl}\Bigg)^2 
\end{eqnarray}
and for $n=3$,
\begin{eqnarray}
&&\left(\frac{1}{9}  \left\langle \hat{q}^6\right\rangle -\frac{2}{3}
  \left\langle \hat{p}^2 \hat{q}^4\right\rangle_{\rm Weyl}+\left\langle \hat{p}^4
    \hat{q}^2\right\rangle_{\rm Weyl}  +\hbar^2   \left\langle
    \hat{q}^2\right\rangle+\hbar^2\left\langle \hat{p}^2\right\rangle
\right)\nonumber\\
&&\times 
 \left(\frac{1 }{9}\left\langle \hat{p}^6\right\rangle-\frac{2
   }{3}\left\langle \hat{p}^4 \hat{q}^2\right\rangle_{\rm Weyl} +
   \left\langle \hat{p}^2 
     \hat{q}^4\right\rangle_{\rm Weyl} +\hbar^2 \left\langle
     \hat{p}^2\right\rangle 
   +\hbar^2  \left\langle \hat{q}^2\right\rangle \right) \nonumber   \\
&&\geq \hbar^2 \Bigg(\frac{\hbar^2}{3}+\frac{1}{2 }\left\langle
  \hat{p}^4\right\rangle +\frac{1}{2}  \left\langle \hat{q}^4\right\rangle
+\left\langle \hat{p}^2 \hat{q}^2\right\rangle_{\rm Weyl}
\Bigg)^2+\Bigg(\frac{1}{3 
}\left\langle \hat{p}^5 \hat{q}\right\rangle_{\rm Weyl}\nonumber\\
&& +\frac{1}{3}  \left\langle
  \hat{p} \hat{q}^5\right\rangle_{\rm Weyl} -\frac{10}{9} \left\langle \hat{p}^3
  \hat{q}^3\right\rangle_{\rm Weyl} \Bigg)^2\,. 
\end{eqnarray}
Except for $n=1$, there is no obvious relationship with minors of the matrices
$M_J'$ introduced in (\ref{MJp}), which were found to be relevant for
eigenstates in our previous analysis.

\subsection{Generalized coherent states}

The saturation property of the harmonic-oscillator ground state, which by
definition satisfies $\hat{a}\psi=0$, is maintained by coherent states defined
by $\sqrt{2\hbar}\hat{a}\psi=\alpha\psi$ with a complex number
$\alpha=\langle\hat{q}\rangle+i\langle\hat{p}\rangle$. Similarly, saturation
properties of higher-order uncertainty relations obeyed by the first $n$
excited states, all subject to the condition can be maintained by generalized
coherent states, for which 
\begin{equation}\label{Coherentn}
 (\sqrt{2\hbar}\,\hat{a})^n\psi=\alpha^n\psi\,. 
\end{equation}
We will first show that these generalized coherent states indeed obey
higher-order uncertainty relations.

As in the case of $\alpha=0$ in the preceding subsection, we introduce two new
operators, $\hat{f}:=(2\hbar)^{n/2}(\hat{a}^n+\hat{a}^{\dagger}{}^n)-\alpha^n$
and $\hat{g}:=(2\hbar)^{n/2}(\hat{a}^n-\hat{a}^{\dagger}{}^n)-\alpha^n$. In a
state $\psi$ that satisfies (\ref{Coherentn}), we again obtain
$\hat{f}\psi=-\hat{g}\psi$ and therefore
\begin{equation} \label{CSn}
  \langle\hat{f}^{\dagger}\hat{f}\rangle \langle
  \hat{g}^{\dagger}\hat{g}\rangle =
  \langle\hat{f}^{\dagger}\hat{g}\rangle \langle 
  \hat{g}^{\dagger}\hat{f}\rangle =
  |\langle\hat{f}^{\dagger}\hat{g}\rangle|^2
\end{equation}
saturating (\ref{cs_saturated}) as before.

The form of these uncertainty relations saturated by a generalized coherent
state depends on the parameter
$\alpha=\langle\hat{q}\rangle+i\langle\hat{p}\rangle$. For instance, for
$n=1$, we do not directly obtain the standard uncertainty relation but rather
compute
\begin{eqnarray}
 \langle \hat{f}^{\dagger}\hat{f}\rangle &=&
 \langle 4\hat{q}^2-2(\alpha+\alpha^*) \hat{q}+|\alpha|^2\rangle = 
4(\Delta
q)^2+\langle\hat{q}\rangle^2+\langle\hat{p}\rangle^2\\ 
 \langle\hat{g}^{\dagger}\hat{g}\rangle &=&
 4(\Delta p)^2+ \langle\hat{q}\rangle^2+\langle\hat{p}\rangle^2\\ 
 \langle\hat{f}^{\dagger}\hat{g}\rangle &=&  4i\langle\hat{q}\hat{p}\rangle-
 2\left(\alpha\langle\hat{q}\rangle+ i\alpha^*\langle\hat{p}\rangle\right)+
 |\alpha|^2 \nonumber\\
&=& i C_{qp}-2\hbar- \langle\hat{q}\rangle^2-\langle\hat{p}\rangle^2
\end{eqnarray}
with the covariance $C_{qp}=\Delta(qp)$. The saturated uncertainty relation
obtained immediately from (\ref{CSn}) then takes the form
\begin{equation}
 (\Delta q)^2(\Delta p)^2-C_{qp}^2+
 \frac{1}{4}
 \left(\langle\hat{q}\rangle^2+\langle\hat{p}\rangle^2\right)\left( (\Delta 
   q)^2+ (\Delta p)^2 -\hbar\right) = \frac{1}{4}\hbar^2\,.
\end{equation}
This equation is equivalent to saturation of the standard uncertainty relation
because $(\Delta q)^2=\hbar/2=(\Delta p)^2$ in a coherent state such that
(\ref{Coherentn}) holds with $n=1$. 

It is possible to evaluate the condition for generalized coherent states
explicitly in terms of energy eigenstates, following the usual procedure for
$n=1$. We will denote these states as $|\alpha,k\rangle$, anticipating the
presence of a second (integer) parameter $k$ because the condition
(\ref{Coherentn}) does not uniquely determine a state for $n>1$ even if
$\alpha$ has been fixed. Using the energy eigenstates $|m\rangle$ as a basis,
we first compute, for integer $0\leq\ell< k$, the inner products
\begin{eqnarray}
 \langle kn+\ell|\alpha,k\rangle &=& \frac{1}{\sqrt{(kn+\ell)!}}
 \left((\hat{a}^{\dagger})^{kn+\ell}|0\rangle\right)^{\dagger}
 |\alpha,k\rangle = \frac{1}{(2\hbar)^{kn/2}}
 \frac{\alpha^{kn}}{\sqrt{(kn+\ell)!}} \langle 
 0|\hat{a}^{\ell}|\alpha,k\rangle \nonumber\\
&=& \frac{\alpha^{kn}}{(2\hbar)^{kn/2}}\frac{\sqrt{\ell!}}{\sqrt{(kn+\ell)!}}
\langle 
 \ell|\alpha,k\rangle =: \alpha^{kn}\frac{\sqrt{\ell!}}{\sqrt{(kn+\ell)!}}
 C_{\ell}
\end{eqnarray}
with $k$ independent constants $C_{\ell}$ (which are
related to one another only by normalization). We then write
\begin{eqnarray}
 |\alpha,k\rangle &=& \sum_{m=0}^{\infty} \langle m|\alpha,k\rangle
 |m\rangle
 = \sum_{\ell=0}^{k-1} C_{\ell} \sqrt{\ell!} \sum_{n=0}^{\infty}
 \frac{\alpha^{kn}}{\sqrt{(kn+\ell)!}} |kn+\ell\rangle\nonumber\\
&=& \sum_{\ell=0}^{k-1} C_{\ell} \frac{\sqrt{\ell!}}{\alpha^{\ell}}
\sum_{n=0}^{\infty} \frac{(\alpha\hat{a}^{\dagger})^{kn+\ell}}{(kn+\ell)!}
|0\rangle\,.
\end{eqnarray}
The infinite series $\sum_{n=0}^{\infty}
(\alpha\hat{a}^{\dagger})^{kn+\ell}/(kn+\ell)!$ in this last expression is
related to the exponential function applied to multiples of
$\alpha\hat{a}^{\dagger}$, but it is not a single such function because $n$ in
the usual series is replaced here by $kn+\ell$. The series encountered here
therefore selects only a subset of the expansion terms of a single exponential
function. Using the basic $k$-th root of unity $u_k=e^{2\pi i/k}$, it is
possible to write our series as a superposition of exponential functions,
\begin{equation}
  \sum_{n=0}^{\infty} \frac{(\alpha\hat{a}^{\dagger})^{kn+\ell}}{(kn+\ell)!}=
  \frac{1}{k} \sum_{j=0}^{k-1} u_k^{-j\ell} \exp(u_k^j\alpha 
\hat{a}^{\dagger})
\end{equation}
in which coefficients have been chosen so as to make unwanted terms cancel
out. Indeed, 
\begin{equation}
 \sum_{j=0}^{k-1} u_k^{-j\ell} \exp(u_k^j\alpha 
\hat{a}^{\dagger})= \sum_{N=0}^{\infty} \frac{1}{N!} u_k^{j(N-\ell)}
(\alpha\hat{a}^{\dagger})^N
\end{equation}
implies the desired equation because
\begin{equation}
 \sum_{j=0}^{k-1} u_k^{j(N-\ell)}= \left\{ \begin{array}{cl} k & \mbox{if
     }N-\ell=kn \mbox{ for some integer }n\\ 0 &
     \mbox{otherwise}\end{array}\right. 
\end{equation}
thanks to the definition of $u_k$.  We can therefore continue our derivation
and write
\begin{equation} \label{alphak}
|\alpha,k\rangle= \sum_{\ell=0}^{k-1} C_{\ell} \frac{\sqrt{\ell!}}{\alpha^{\ell}}
\frac{1}{k} \sum_{j=0}^{k-1} u_k^{-jl} \exp(u_k^j\alpha
\hat{a}^{\dagger})|0\rangle
= \frac{1}{k} e^{\frac{1}{2}|\alpha|^2} \sum_{j=0}^{k-1} D_j
|u_k^j\alpha\rangle
\end{equation}
with the standard coherent states $|\beta\rangle= e^{-\frac{1}{2}|\beta|^2}
\exp(\beta\hat{a}^{\dagger}) |0\rangle$ and new constants
\begin{equation}
 D_j= \sum_{\ell=0}^{k-1} \frac{\sqrt{\ell!}}{\alpha^{\ell}} u_k^{-j\ell}
 C_{\ell}\,.
\end{equation}

Multiplying the parameter
$\alpha=\langle\hat{q}\rangle+i\langle\hat{p}\rangle$ of a standard coherent
state with a power of a basic root of unity $u_k$ in the superposed coherent
states $|u_k^j\alpha\rangle$ of (\ref{alphak}) rotates the peak position
$(\langle\hat{q}\rangle,\langle\hat{p}\rangle)$ in phase space by a multiple
of a fixed angle $2\pi/k$.  According to (\ref{alphak}), a generalized
coherent state $|\alpha,k\rangle$ is therefore a superposition of $k$ standard
coherent states with peaks $(\langle\hat{q}\rangle,\langle\hat{p}\rangle)$
placed at equal distances on a circle of radius $|\alpha|$. The $k$-th
eigenstate of the harmonic oscillator is the limit in which these peaks
approach one another at the center, for suitable $C_{\ell}$. Using
\cite{GenCoh3}, these generalized coherent states are the same as those
introduced by Titulaer and Glauber in \cite{GenCoh}; see also
\cite{GenCoh2}. However, to the best of our knowledge, the relation to
saturated uncertainty relations and energy eigenstates is new.

\section{Anharmonic oscillators}

We now demonstrate that the methods developed in
Section~\ref{sec:eigs_and_mnt} can be used to find perturbed eigenvalues for
an anharmonic oscillator. Here we take $H = \frac{1}{2}(q^2 + p^2) + \epsilon
q^4$. 

\subsection{Moment method}

Using the same techniques as for the harmonic oscillator (but now
setting $\hbar=1$), we obtain the following recurrence relations for the
moments:
\begin{eqnarray}\label{anharm_recur1}
T_{m+2,n} &+& T_{m,n+2} - \frac{n(n-1)}{4}T_{m,n-2} -
\frac{m(m-1)}{4}T_{m-2,n} - 2\lambda T_{m,n} \\ \nonumber 
&+&\epsilon \left( 2\hat{T}_{m+4,n} - 3n(n-1)T_{m+2,n-2} +
  \frac{1}{8}n(n-1)(n-2)(n-3)T_{m,n-4} \right)= 0 
\end{eqnarray}
and
\begin{equation} \label{anharm_recur2}
m\hat{T}_{m-1,n+1} = n\hat{T}_{m+1,n-1} + \epsilon\left( 4n\hat{T}_{m+3,n-1} -
  n(n-1)(n-2)T_{m+1,n-3}\right) \,.
\end{equation}

Setting $n=0$ in (\ref{anharm_recur1}) and $n=1$ in (\ref{anharm_recur2})
while shifting $m$ to $m+1$, and combining to eliminate $T_{m,2}$ gives
\begin{equation}\label{anharm_recur3}
\frac{(m+2)}{(m+1)}T_{m+2,0} - 2 \lambda T_{m,0} - \frac{m(m-1)}{4} T_{m-2,0}
+ 2\epsilon \frac{(m+3)}{(m+1)}T_{m+4,0} = 0 \,.
\end{equation}
Then using (\ref{anharm_recur2}) with $n$ shifted to$n+1$ and $m$ to $m-1$
results in
\begin{equation} \label{anharm_recur4}
T_{m-2,n+2} =\frac{(n+1)}{(m-1)} T_{m,n} + \epsilon \left(
  4\frac{(n+1)}{(m-1)}T_{m+2,n} - \frac{(n+1)(n)(n-1)}{(m-1)}
  T_{m,n-2}\right)\,. 
\end{equation}
We now assume an expansion for the moments in powers of $\epsilon$
\begin{equation}
T_{m,n} = \sum_{k} T^{(k)}_{m,n} \epsilon^k
\end{equation}
and similarly for the eigenvalues,
\begin{equation}\label{anharm_lambda_expansion}
\lambda = \sum_k \lambda_{(k)} \epsilon^k.
\end{equation}
Using Equations~(\ref{anharm_recur3})--(\ref{anharm_lambda_expansion}), we can
solve order by order for the moments in terms of the $\lambda_{(k)}$.

For the odd moments, we first note that, at zeroth order, all of them are zero
(as we know well from the harmonic oscillator): 
\begin{equation}
 T_{\textrm{\scriptsize
    odd},\textrm{\scriptsize odd}}^{(0)}=T_{\textrm{\scriptsize
    odd},\textrm{\scriptsize even}}^{(0)}=T_{\textrm{\scriptsize
    even},\textrm{\scriptsize odd}}^{(0)}=0\,.
\end{equation}
Then setting $m=0$ and $n=1$ in (\ref{anharm_recur2}) gives $T_{1,0}^{(1)} =
0$. Using this and (\ref{anharm_recur3}) with $m$ odd gives
$T_{\textrm{\scriptsize odd},0}^{(1)}=0$. Taking $n=0$ in
(\ref{anharm_recur2}) gives $T_{m,1} = 0$ at all orders in
$\epsilon$. Combining these two results with (\ref{anharm_recur4}) implies
that the rest of the odd moments vanish: 
\begin{equation}
 T_{\textrm{\scriptsize
    odd},\textrm{\scriptsize odd}}^{(1)}=T_{\textrm{\scriptsize
    odd},\textrm{\scriptsize even}}^{(1)}=T_{\textrm{\scriptsize
    even},\textrm{\scriptsize odd}}^{(1)}=0\,.
\end{equation}
We can apply this argument repeatedly to find that the odd moments vanish at
all orders in $\epsilon$.

Using the recurrence relations following the procedure detailed in
Section~\ref{sec:eigs_and_mnt}, we find to first order in $\epsilon$
\begin{eqnarray}
\det{(A_1')} &=& \left(\lambda_{(0)}-\frac{1}{2}\right)
\left(\lambda_{(0)}+\frac{1}{2}\right)\\ 
&&- \epsilon \left(\frac{1}{4}\lambda_{(0)}   \left(12 \lambda_{(0)}^2-8
    \lambda_{(1)}+3\right) \right) + O(\epsilon^2)\\ 
\det(A_2') &=& \frac{1}{4} \left(\lambda_{(0)}-\frac{3}{2}\right)
\left(\lambda_{(0)}-\frac{1}{2}\right) \left(\lambda_{(0)}+\frac{1}{2}\right)
\left(\lambda_{(0)}+\frac{3}{2}\right)\\ 
& &-\epsilon \left(\frac{1}{32}  \lambda_{(0)} \left(80 \lambda_{(0)}^4-32
    (\lambda_{(1)}+4) \lambda_{(0)}^2+40 \lambda_{(1)}+3\right)\right) +
O(\epsilon^2)\,. 
\end{eqnarray}

At zeroth order in $\epsilon$, we recover our results for the harmonic
oscillator. Setting $\lambda_{(0)} = 1/2$, we find:
\begin{eqnarray}
\det{(A_1')} &=&\epsilon  \left(\lambda_{(1)}-\frac{3}{4}\right) +
O(\epsilon^2)\\ 
\det{(A_2')} &=&  \epsilon  \left(\frac{3}{8} - \frac{1}{2}
  \lambda_{(1)}\right) + O(\epsilon^2)\,. 
\end{eqnarray}
Positivity of these determinants then yields $\lambda_{(1)} \ge 3/4$ and
$\lambda_{(1)} \le 3/4$. Hence, $\lambda_{(1)} = 3/4$. Performing the same
process with $\det(A_2')$ and $\det(A_3')$ using $\lambda_{(0)} = 3/2$ yields
$\lambda_{(1)} = 15/4$.
Thus we have:
\begin{eqnarray}
E_0 &=& \frac{1}{2} + \frac{3}{4}\epsilon + O(\epsilon^2)\\
E_1 &=& \frac{3}{2} + \frac{15}{4}\epsilon+ O(\epsilon^2)
\end{eqnarray}
in agreement with the results from ordinary perturbation theory.

Note that at first order in $\epsilon$, the energy eigenstates saturate the
inequalities just as they did for the harmonic oscillator. Computations at
higher order indicate that similar saturation results hold at each order in
perturbation theory, although for higher orders in $\epsilon$, one must go to
higher $n$ in order for $\det(A_n')\ge 0$ to be saturated.

\subsection{Commutator method}

An alternative route to perturbated eigenvalues, which may sometimes be more
feasible, proceeds by applying suitable commutator relationships.  Following
\cite{TransMom}, we can derive recurrence relations for moments of energy
eigenstates: We have $\langle n|[\hat{H},\hat{W}]|n\rangle=0$ for any operator
$\hat{W}$, with eigenstates $|n\rangle$ of
$\hat{H}=\frac{1}{2}m^{-1}\hat{p}^2+V(\hat{q})$. Choosing
$\hat{W}_1=\hat{q}^{k-2}$ and $\hat{W}_2=\hat{q}^{k-1}\hat{p}$, respectively,
for some fixed $k$, we obtain
\begin{eqnarray}
 [\hat{H},\hat{W}_1] &=& -i\hbar \frac{k-2}{m} \hat{q}^{k-3}\hat{p}- \hbar^2
 \frac{(k-2)(k-3)}{2m} \hat{q}^{k-4} \label{W1}\\
\mbox{} [\hat{H},\hat{W}_2] &=& -2i\hbar(k-1) \hat{q}^{k-2}(\hat{H}-V(\hat{q}))-
 \hbar^2 \frac{(k-1)(k-2)}{2m} \hat{q}^{k-3}\hat{p}+ i\hbar
 \hat{q}^{k-1}V'(\hat{q})\,. \label{W2}
\end{eqnarray}
We combine these two equations (set equal to zero) and (divided by
$i\hbar$) write
\begin{equation}
 0=-2(k-1)E_n \langle \hat{q}^{k-2}\rangle_n+ 2 (k-1) \langle
 \hat{q}^{k-2}V(\hat{q})\rangle_n- \hbar^2 \frac{(k-1)(k-2)(k-3)}{4m}
 \langle\hat{q}^{k-4}\rangle_n+
 \langle\hat{q}^{k-1}V'(\hat{q})\rangle_n\,.
\end{equation} 

For a quartic anharmonicity, such that $V(q)=\frac{1}{2}m\omega^2 q^2+\epsilon
q^4$, we have
\begin{equation}
 0=-2(k-1) E_n \langle \hat{q}^{k-2}\rangle_n- (k-1)(k-2)(k-3)
 \frac{\hbar^2}{4m} \langle\hat{q}^{k-4}\rangle_n+
 m\omega^2k\langle\hat{x}^k\rangle_n+ 2\epsilon
 (k+1)\langle\hat{q}^{k+2}\rangle_n\,.
\end{equation}
Starting with $k=1$, the first four recurrence steps are:
\begin{eqnarray}
 0&=&
 m\omega^2\langle\hat{q}\rangle_n+4\epsilon\langle\hat{q}^3\rangle_n \label{Rec1}
 \\
 0&=& -2E_n
 +2m\omega^2\langle\hat{q}^2\rangle_n+6\epsilon\langle\hat{q}^4\rangle_n 
\label{Rec2}\\ 
0&=& -4E_n\langle\hat{q}\rangle_n+3m\omega^2\langle\hat{q}^3\rangle_n+
8\epsilon\langle\hat{q}^5\rangle_n \label{Rec3}\\
0&=& -6E_n\langle\hat{q}^2\rangle_n- \frac{3\hbar^2}{2m}+ 4m\omega^2
\langle\hat{q}^4\rangle_n+10\epsilon \langle\hat{q}^6\rangle_n\,. \label{Rec4}
\end{eqnarray}
Assuming $\epsilon$ to be small and expanding $\langle\hat{q}^k\rangle_n=
\sum_{j=0}^{\infty}\langle\hat{q}^k\rangle_{n,j}\epsilon^j$, we have
$\langle\hat{q}\rangle_{n,0}=0$ from (\ref{Rec1}), which implies
$\langle\hat{q}^3\rangle_{n,0}=0$ from (\ref{Rec3}), such that
$\langle\hat{q}\rangle_{n,1}=0$ from (\ref{Rec1}).

For even powers, $\langle\hat{q}^2\rangle_{n,0}=E_n/m\omega^2$ from
(\ref{Rec2}) and $\langle\hat{q}^4\rangle_{n,0}=\frac{3}{2}E_n^2/m^2\omega^4+
\frac{3}{8} \hbar^2/m^2\omega^2$ from (\ref{Rec4}). This value then appears in
$\langle\hat{q}^2\rangle_{n,1}=-3\langle\hat{x}^4\rangle_{n,0}/m\omega^2$ from
(\ref{Rec2}).
We obtain some of the moments including $\hat{p}$ from (\ref{W1}) and
(\ref{W2}). Setting $k=4$ in (\ref{W1}) shows that
$\langle\hat{q}\hat{p}+\hat{p}\hat{q}\rangle_n=0$ in all energy
eigenstates. Setting $k=2$ in (\ref{W2}) and {\em not} using
$\hat{H}|n\rangle=E_n$ implies
\begin{equation}
 \langle\hat{p}^2\rangle_n= m\langle\hat{q}V'(\hat{q})\rangle_n=
 m^2\omega^2\langle\hat{q}^2\rangle_n+ 4m\epsilon\langle\hat{q}^4\rangle_n\,,
\end{equation}
the final equality for our anharmonic oscillator. Using the results for low
orders of $q$-moments, we have
\begin{eqnarray}
 \langle\hat{p}^2\rangle_{n,0} &=& m^2\omega^2\langle\hat{q}^2\rangle_{n,0}=
 mE_n\\
\langle\hat{p}^2\rangle_{n,1} &=& m^2\omega^2\langle\hat{q}^2\rangle_{n,1}+
4m\langle\hat{q}^4\rangle_{n,0}= m\langle\hat{q}^4\rangle_{n,0}\,.
\end{eqnarray}

To first order in $\epsilon$, we therefore compute
\begin{eqnarray}
 \langle\hat{q}^2\rangle_n &=& \langle\hat{q}^2\rangle_{n,0}+\epsilon
 \langle\hat{q}^2\rangle_{n,1} +O(\epsilon^2) = \frac{E_n}{m\omega^2}-
 \frac{9\epsilon}{8m^3\omega^6} (4E_n^2+\hbar^2\omega^2)+O(\epsilon^2)\\
 \langle\hat{p}^2\rangle_n &=& \langle\hat{p}^2\rangle_{n,0}+\epsilon
 \langle\hat{p}^2\rangle_{n,1} +O(\epsilon^2) = mE_n+
 \frac{3\epsilon}{8m\omega^4} (4E_n^2+\hbar^2\omega^2)+O(\epsilon^2)\,.
\end{eqnarray}
The uncertainty relation implies
\begin{equation}
 \langle\hat{q}^2\rangle_n\langle\hat{p}^2\rangle_n= \frac{E_n^2}{\omega^2}-
 \frac{3\epsilon E_n}{4m^2\omega^6} (4E_n^2+\hbar^2\omega^2)+O(\epsilon^2)\geq
 \frac{\hbar^2}{4}\,.
\end{equation}
At zeroth order in $\epsilon$, this implies $E_n\geq
\frac{1}{2}\hbar\omega$. If we use an $\epsilon$-expansion of
$E_n=\sum_{j=0}^{\infty} E_{n,j}\epsilon^j$ at this stage, we obtain
\begin{equation}
 E_n\geq\frac{1}{2}\hbar\omega+ \frac{3}{4}
 \frac{\epsilon\hbar^2}{m^2\omega^2}+O(\epsilon^2)\,.
\end{equation}

The present formulas indicate that neither the moments nor the uncertainty
relations and bounds on eigenvalues are analytic in $\omega$, such that we
cannot take a $\omega\to0$ limit for a single quartic potential.

\section{Discussion}

We have presented a new method that allowed us to rederive known results about
energy eigenvalues using only properties of the algebra of observables. The
results are therefore representation-independent, and the method can be
applied to systems that do not have a Hilbert-space representation mainly
owing to violations of associativity. Even in standard, associative quantum
mechanics, we have been able to derive new results related to how excited
states saturate higher-order uncertainty relations and connections between
excited states and generalized coherent states.

Our new method starts with the algebraic definition
(\ref{h_constraint}), or 
\begin{equation}\label{Aspectrum}
\langle\hat{A}(\hat{H} - \lambda {\mathbb I})
\rangle_{\lambda} = 0\,,
\end{equation} 
of an eigenstate $|\rangle_{\lambda}$ with eigenvalue $\lambda$, which has to
be satisfied for all algebra elements $\hat{A}$. In particular, the definition
is taylored to strict eigenstates which are normalizable since
$\langle{\mathbb I}\rangle_{\lambda}$ must be finite for the equation to be
meaningful for all $\hat{A}$. The method can therefore be used only for
eigenvalues in the discrete part of the spectrum of $\hat{H}$.

If we try to work out the algebraic conditions for eigenstates in simple cases
which are known to imply continuous spectra, we can easily find
inconsistencies. For instance, taking $\hat{H}=\hat{p}$ as the momentum
operator of a particle on the real line and $\hat{A}=\hat{q}$ in
(\ref{Aspectrum}), we obtain the equation
\begin{equation}
 {\rm Im}\langle
 \hat{q}(\hat{p}-\lambda {\mathbb I})\rangle= \frac{1}{2i}\langle
 [\hat{q},\hat{p}]\rangle =\frac{1}{2}\hbar
\end{equation}
while the eigenvalue condition for $\lambda$ would require the left-hand side
to equal zero.

For the free-particle Hamiltonian, $\hat{H}=\hat{p}^2$, we obtain
$\langle\hat{p}^2\rangle-\lambda=0$ from (\ref{Aspectrum}) with
$\hat{A}={\mathbb I}$, and 
 \begin{equation}
 {\rm Im}\langle
 \hat{q}\hat{p}(\hat{p}^2-\lambda {\mathbb I})\rangle= \frac{1}{2i}\langle
 [\hat{q},\hat{p}^3]-\lambda[\hat{q},\hat{p}]\rangle
 =\frac{1}{2}\hbar(3\langle\hat{p}^2\rangle-\lambda)=0 
\end{equation}
from $\hat{A}=\hat{q}\hat{p}$. Combining these two equations, only $\lambda=0$
is allowed. 
However, 
\begin{equation}
 {\rm Im}\langle
 \hat{q}(\hat{p}^2-\lambda {\mathbb I})\rangle= \frac{1}{2i}\langle
 [\hat{q},\hat{p}^2]\rangle
 =\hbar\langle\hat{p}\rangle=0
\end{equation}
then implies $(\Delta p)^2=0$, which is inconsistent with  Heisenberg's
uncertainty relation.

In some cases, the range of eigenvalues in the spectrum according to
(\ref{Aspectrum}) may nevertheless be continuous. Because the algebraic
condition for the spectrum is representation independent, an algebra that can
be represented on a non-separable Hilbert space may lead to a continuous set
of eigenvalues even for normalizable eigenstates. As an example, consider a
particle moving on a circle. The corresponding algebra can be generated by
three basic operators, $\hat{p}$, $\hat{S}$ and $\hat{C}$, with relations
$[\hat{p},\hat{S}]=-i\hbar\hat{C}$, $[\hat{p},\hat{C}]=i\hbar\hat{S}$ and
$[\hat{C},\hat{S}]=0$. (The operators $\hat{S}$ and $\hat{C}$ quantize the
sine and cosine of the angle.) This linear algebra has the Casimir element
$\hat{K}=\hat{S}^2+\hat{C}^2$ which we may require to equal $\hat{K}={\mathbb
  I}$ as a further relation in the generated algebra. Our Hamiltonian is
$\hat{H}=\hat{p}$.

The condition $\langle\hat{p}^{n-1}(\hat{H}-\lambda)\rangle=0$ for $n\geq 1$
implies that $\langle\hat{p}^n\rangle=\lambda^n=\langle\hat{p}\rangle^n$, and
therefore all central $p$-moments
$\langle(\hat{p}-\langle\hat{p}\rangle)^n\rangle=0$ vanish. More generally, it
follows that $\langle\hat{A}(\hat{p}-\langle\hat{p}\rangle)\rangle=
\langle\hat{A}(\hat{H}-\lambda)\rangle=0$ for all $\hat{A}$. All generalized
uncertainty relations are therefore identically satisfied because the lower
bound in the Cauchy--Schwarz inequality is always zero for eigenstates. For
any real $\lambda$, there is therefore an eigenstate with this eigenvalue.

This result is in agreement with Hilbert-space representations of the algebra,
which are not unique up to unitary equivalence. Inequivalent representations
are labeled by a real number $0\leq \epsilon<1$, such that the momentum
spectrum for a given $\epsilon$ is ${\mathbb Z}+\epsilon$. The direct sum of
all inequivalent representations is non-separable. If the representation is
not fixed, all real numbers can be realized as eigenvalues of $\hat{p}$.

As these examples demonstrate, the spectrum cannot always be fully analyzed
based on the algebraic condition (\ref{Aspectrum}), unless it is strictly
discrete. As a consequence, it remains an open question how the continuous
spectrum could be defined in non-associative quantum mechanics.

\section*{Acknowledgements}

This work was supported in part by NSF grant PHY-1912168.


\end{document}